\documentclass[sigconf]{acmart}

\usepackage{booktabs} 

\usepackage[normalem]{ulem}
\usepackage{amsmath,amsthm,amssymb}
\usepackage{graphicx}
\usepackage{algorithm}
\usepackage[noend]{algpseudocode}
\usepackage{verbatim}
\usepackage{adjustbox}
\usepackage{xpatch,amsthm}
\usepackage{mathtools}
\usepackage{subcaption}
\usepackage{enumitem}
\setitemize{noitemsep,topsep=0pt,parsep=0pt,partopsep=0pt}
\usepackage{soul}
\usepackage{xcolor}

\setcopyright{acmcopyright}
\begin{document}

%

\title{Slack Squeeze Coded Computing for Adaptive Straggler Mitigation}
\author[K. Narra]{Krishna Giri Narra}
\authornote{Both authors contributed equally to this research.}
\affiliation{%
  \institution{Department of Electrical Engineering, University of Southern California}
  \city{Los Angeles}
  \state{CA}
  \country{USA}
}
\author[Z. Lin]{Zhifeng Lin}
\authornotemark[1]
\affiliation{%
  \institution{Department of Electrical Engineering, University of Southern California}
  \city{Los Angeles}
  \state{CA}
  \country{USA}
}
\author[M. Kiamari]{Mehrdad Kiamari}
\affiliation{%
  \institution{Department of Electrical Engineering, University of Southern California}
  \city{Los Angeles}
  \state{CA}
  \country{USA}
}
\author[S. Avestimehr]{Salman Avestimehr}
\affiliation{%
  \institution{Department of Electrical Engineering, University of Southern California}
  \city{Los Angeles}
  \state{CA}
  \country{USA}
}
\author[M. Annavaram]{Murali Annavaram}
\affiliation{%
  \institution{Department of Electrical Engineering, University of Southern California}
  \city{Los Angeles}
  \state{CA}
  \country{USA}
}

\begin{abstract}

While performing distributed computations in today's cloud-based platforms, execution speed variations among compute nodes can significantly reduce the performance and create bottlenecks like stragglers. Coded computation techniques leverage coding theory to inject computational redundancy and mitigate stragglers in distributed computations. In this paper, we propose a dynamic workload distribution strategy for coded computation called Slack Squeeze Coded Computation ($S^2C^2$). $S^2C^2$ squeezes the compute slack (i.e., overhead) that is built into the coded computing frameworks by efficiently assigning work for all fast and slow nodes according to their speeds and without needing to re-distribute data. We implement an LSTM-based speed prediction algorithm to predict speeds of compute nodes. We evaluate $S^2C^2$ on linear algebraic algorithms, gradient descent, graph ranking, and graph filtering algorithms. We demonstrate 19\% to 39\% reduction in total computation latency using $S^2C^2$ compared to job replication and coded computation. We further show how $S^2C^2$ can be applied beyond matrix-vector multiplication.
\end{abstract}

\copyrightyear{2019} 
\acmYear{2019} 
\acmConference[SC '19]{The International Conference for High Performance Computing, Networking, Storage, and Analysis}{November 17--22, 2019}{Denver, CO, USA}
\acmBooktitle{The International Conference for High Performance Computing, Networking, Storage, and Analysis (SC '19), November 17--22, 2019, Denver, CO, USA}
\acmPrice{15.00}
\acmDOI{10.1145/3295500.3356170}
\acmISBN{978-1-4503-6229-0/19/11}

%
%


%

\maketitle


\section{Introduction}

Cloud computing and large distributed frameworks like Apache Spark~\cite{spark} are being widely used as they enable the execution of large-scale applications, such as machine learning and graph analytics on data sizes of the order of tens of terabytes and more, efficiently and at lower cost. However, as we ``scale out'' computations across many distributed nodes, one needs to deal with the ``system noise'' that is due to several factors such as heterogeneity of hardware, hardware failures, disk IO delay, communication delay, operating system issues, maintenance activities, and power limits~\cite{ananthanarayanan2010reining}. 
System noise leads to uneven execution latencies where different servers may take different amount of time to execute the same task, even if the servers have identical hardware configuration. In the extreme case, a server may be even an order of magnitude slower than the remaining servers, which we refer to as a straggler node. Such speed variations create significant delays in task executions and can also lead to major performance bottlenecks, since the master node waits for the slowest worker to finish its task. This phenomenon results in tail latency which can be defined as the high percentile completion latency of the distributed tasks. If the number of servers within a cluster experiencing this speed variance increases, the probability of having long tail latency increases exponentially \cite{tail}.

Replication approaches are commonly used today to deal with the straggler delay bottleneck. For example, in distributed computation and storage frameworks like Hadoop MapReduce~\cite{hadoop} and Spark, data that needs to be processed is split into partitions by the master node, and each data partition is replicated across a subset of  workers. The master node then keeps track of task progress on each worker node. After completion of a certain fraction of the tasks, if the master observes that a particular worker node's progress is slow, it schedules a copy of that slow task to be executed on another node which contains a copy of the data partition to be processed. Whichever node, either the original or the replicated node, finishes first will return the results to the master node and the other copy of the task is terminated. This technique is used in the Hadoop MapReduce framework \cite{hadoop}. However this technique is reactive since the master node waits until most of the tasks finish their execution before launching a replica of tasks running on straggler nodes. In addition the replica can only be launched on a restrictive subset of nodes that have a data copy. This process significantly degrades the overall execution time. 

Recently, it has been been shown that ``coding'' can provide a novel approach to mitigate the tail latency caused by straggler nodes, and a new framework named ``coded computation'' is proposed~\cite{speedUpML,LMA16_unify,reisizadehmobarakeh2017coded,polyCodes,dutta2016short}. The key idea of coded computation frameworks is to inject computation redundancy in an unorthodox coded form (as opposed to the state-of-the-art replication approaches) in order to create robustness to stragglers. For example, it was shown in prior work~\cite{speedUpML} that error correcting codes (e.g.,  Maximum-Distance-Separable (MDS) codes~\footnote{MDS codes are an important class of block codes since they have the greatest error correcting and detecting capabilities. For more information see~\cite{hill1986first} Chapter 16.}) can be utilized to create redundant computation tasks for linear computations (e.g., matrix multiplication). 

\textbf{Overview of MDS coding:} An $(n,k)$-MDS coded computation first decomposes the overall computation into $k$ smaller tasks, for some $k \leq n$. Then it encodes them into $n$ coded tasks using an $(n,k)$-MDS code, and assigns each of them to a node to compute. From the desirable ``any $k$ of $n$'' property of the MDS code, the overall computation can be completed once the results from the \emph{fastest} $k$ out of $n$ coded tasks complete, without waiting for the tasks still running on the slow nodes (or stragglers). Coded computation schemes use these $(n,k)$ parameters to encode the data and to determine how much of the coded dataset is processed by each of the $n$ compute nodes. The smaller the $k$ value (i.e. the more conservative, highly redundant), the larger the amount of computation performed by each node in the cluster. 

The end user must decide on the value of $n$ and $k$ at the application launch time so as to determine the data encoding and decoding process. For $(n,k)$-MDS code the assumption is that there may be at most $n-k$ very slow nodes or failures. But estimating the number of stragglers during the application launch  time is a challenging task~\cite{clones,late}. As such application designers may assume  worst case scenarios for the number of stragglers by specifying a conservative $n-k$. 

If the number of persistent stragglers during a particular execution instance is fewer than what the coding scheme is built to support, efficiency of coded computing drops. For instance, in MDS coding as explained above there is no significant performance benefit if there are fewer than $n-k$ stragglers, since the coded computation still has to wait for $k$ nodes to complete their execution. In  cloud  computing systems partial stragglers are more often encountered i.e.,  nodes  that  are  slower  but  can  do  partial amount of work assigned to them. Existing coded computation schemes always waste the compute capability of the $n-k$ partial stragglers and do not take advantage of the fact that data which is needed for computation already exists with them and they can do partial amount of useful work (More on this in section \ref{real_deploy}). It is this lack of elasticity that makes coded computing unpalatable in large scale cluster settings.  What is ideal is to allow the developer to select high redundancy coding to be conservative (essentially assuming a reasonable worst case straggler scenario) but allow a workload scheduler to decide how much redundant computing to perform based on observed speed variations in a distributed or cloud computing environment. 

In this work, we design a new dynamic workload distribution  strategy for coded computing that is elastic with the speeds of nodes measured during runtime, irrespective of how much redundancy is chosen for creating the coded data. Our proposed $S^2C^2$ (Slack Squeeze Coded Computing) strategy adapts to varying number of stragglers by squeezing out any computation slack that may be built into the coded computation to tolerate the worst case execution scenarios. The performance of $S^2C^2$ is determined by the actual speeds measured and actual number of straggler nodes seen rather than by the redundancy used in encoding the data. As the speeds of nodes change, $S^2C^2$ responds by appropriately increasing or decreasing the amount of work allocated to each node in the cluster to maximize the performance.

To predict the speeds of the nodes as they change during runtime we use a prediction mechanism. We model speed prediction into a time series forecasting problem and use a Long Short-Term Memory (LSTM) based learning model to predict the speeds. These predicted speeds are used by $S^2C^2$ to do work allocation among the nodes.

In summary, the main contributions in this paper are as follows:
\begin{itemize}
\item We empirically measure the speed variations of compute nodes in a large-scale cloud computing cluster. Using the measured data we design an LSTM based model to predict the speed of each node in the next computation epoch.

\item We propose $S^2C^2$ which exploits the data redundancy available in coded data and elastically distributes work based on speeds predicted from the LSTM model. $S^2C^2$  increases performance without compromising on robustness. We also propose a new fine-grained replication baseline that combines over-decomposition of data \cite{CharmppOOPSLA93} and speed prediction based workload distribution.

\item We propose two variations of $S^2C^2$ and evaluate their performance on our local cluster and on a commercial cloud setting while running machine learning and graph ranking workloads.

\item While executing algorithms such as  gradient descent, graph ranking, and graph filtering  $S^2C^2$ is able to reduce the total compute latency by up to $39.3$\% over the conventional coded computation and by up to $19\%$ over the fine-grained replication baseline.

\item Finally, we go beyond matrix vector multiplication to demonstrate the versatility of $S^2C^2$ by applying its workload distribution and scheduling strategies on top of polynomial code~\cite{polyCodes}, a coded computing strategy for polynomial computations.
 
\end{itemize}

Rest of the paper is organized as follows: section 2 provides  background on coded computation, section 3 describes speed prediction and overheads of coded computation, section 4 describes proposed $S^2C^2$ algorithm, section 5 describes extensions to non-linear coded computing, section 6 provides implementation and system details, section 7 shows evaluations, section 8 describes related work.

\section{Coded Computing Background}\label{background}
In this section we briefly introduce the coded computation. For clarity of explanation we first focus on how coded computing is applied for linear algebraic operations.
Let us consider a  distributed matrix multiplication problem 
where a master node wants to multiply a matrix ${\bf A}$ with the input vector ${\overrightarrow x}$ to compute ${\bf A}{\overrightarrow x}$. The data matrix ${\bf A}$ is distributed across $3$ worker nodes
on which the matrix multiplication will be executed in a distributed manner.

One natural approach to tackle this problem is to vertically and evenly divide the data matrix ${\bf A}$ 
into $3$ sub-matrices, each of which is stored on one node. Then when each node receives the input ${\overrightarrow x}$, it simply multiplies its locally stored sub-matrix with ${\overrightarrow x}$ and returns the results, and the master vertically concatenates the returned matrices to obtain the final result. However,  since uncoded approach relies on successfully retrieving the task results from \emph{all} $3$ nodes, it has a major drawback that once one of the nodes runs slow, the computation may take long to finish. Coded computation framework deals with slow or straggler nodes by optimally creating redundant computation tasks. An MDS-coded computing scheme vertically partitions the data matrix ${\bf A}$ into $2$ sub-matrices ${\bf A}_1$ and ${\bf A}_2$, and creates one redundant task by summing ${\bf A}_1$ and ${\bf A}_2$. Then ${\bf A}_1$, ${\bf A}_2$ and ${\bf A}_1+{\bf A}_2$ are stored on worker nodes 1, 2, and 3 
respectively. Each node then performs computations on its own data matrix partition. 
In this case the final result is obtained once the master receives the task results from any 2 out of the 3 nodes, without needing to wait for the slow/straggler node. Let us assume worker node 2 is a straggler and the master node only collects results from node 1 and 3. Then the master node can compute ${\bf A}_2\overrightarrow{x}$ by subtracting the computed task of node 1, i.e. ${\bf A}_1\overrightarrow{x}$, from the computed task of node 3, i.e. $({\bf A}_1+{\bf A}_2)\overrightarrow{x}$.

\textbf{Broader use of coded computing:} MDS-coded computing can inject redundancy to tolerate stragglers in linear computations.  Coded computing is applicable to a wider range of compute intensive algorithms, going beyond linear computations. Polynomial coded computing~\cite{polyCodes} can tolerate stragglers in bilinear computations such as Hessian matrix computation. Lagrange coded computing~\cite{yu2018lagrange} can add coded redundancy to tolerate stragglers in any arbitrary multivariate polynomial computations such as general tensor algebraic functions, inner product functions, function computing outer products, and tensor contractions~\cite{renteln2013manifolds}. Recent works ~\cite{learnCode, narra2019collage, kosaian2019parity} demonstrate promising results by extending coded computing to non-linear applications such as deep learning inference. Finally, coded computing has also been leveraged recently for secure and privacy preserving distributed machine learning~\cite{CPML}.

\section{Motivation}\label{motivation}

\subsection{Straggler Mitigation Overheads} \label{subsec:overheads}

Consider an uncoded strategy with $r$-replication i.e., each data partition is replicated across $r$ different worker nodes where $r$ is the replication factor. Consider a node $N$ executing task $T$ on data partition $DP_T$. If the node $N$ is determined to be a straggler at some future time, the master node can replicate task $T$ on any one of the nodes which has a replica of data partition $DP_T$ to speed it up. However, there are two challenges. First, when should the master determine that $N$ is a straggler? Second, even if the master has early knowledge of $N$ as a straggler, it is restricted to launching the task $T$ only on a subset of nodes that have the required data partition $DP_T$. Third, in the worst case if all the nodes with replicas are also stragglers i.e., if the system has $r$ stragglers, the uncoded replication strategy cannot speed up computation at all. An alternative is to move the data partition $DP_T$ to another available faster node and execute $T$ on that node. This option forces data transfer time into the critical path of the overall execution time.

Next let us consider the $(n,k)$-MDS coded computation on matrix multiplication. The master node divides the original matrix $A$ into $k$ sub-matrices, encodes them into $n$ partitions and distributes them to workers. As we discussed before, a small $k$ needs to be chosen for dealing with worst case scenarios. However, this over-provisioning with a small $k$ comes with a price. If the original data size is $S$, then each of the worker nodes must compute on a coded partition of size $(S/k)$. If $k$ becomes smaller, each worker node has to execute a larger fraction of the computation independent of their actual speeds. On the other hand with a large $k$ the robustness of the computation decreases. This is a difficult tradeoff since the selection of $k$ must be done prior to creating a correct encoding and decoding strategy, and distribution of the encoded data partitions appropriately to all nodes, which are usually done once  before executing the given workload. 

One solution to deal with the straggler uncertainty with MDS-coded computation is to store multiple encoded partitions in each worker node, such that the system can \textit{adapt} and choose the appropriate encoded partition dynamically when the number of stragglers changes in the cluster. For example, in a cluster with 12 worker nodes, each worker node can store a $(12,9)$-MDS encoded partition and a  $(12,10)$-MDS encoded partition at the same time. Assume the original data size is $S$, when it's observed there are three straggling nodes, $(12,9)$-MDS-coded computation will be performed with each worker node operating on an encoded partition of size $(S/9)$; and when it's observed there are fewer straggling nodes, $(12,10)$-MDS-coded computation is  performed with each worker node operating on partition of size $(S/10)$. This approach is optimal only for two scenarios, and supporting a wider range of scenarios means storing more copies of the encoded data. This dramatically increases the storage overhead. It is possible to encode the data at run time and redistribute the large data partitions based on measured speeds and slow node count. However, this will dramatically increase the communication overhead and is not practical.


\begin{figure}
    \centering
    \includegraphics[width=\linewidth, height=4cm, trim=4 4 4 4, clip]{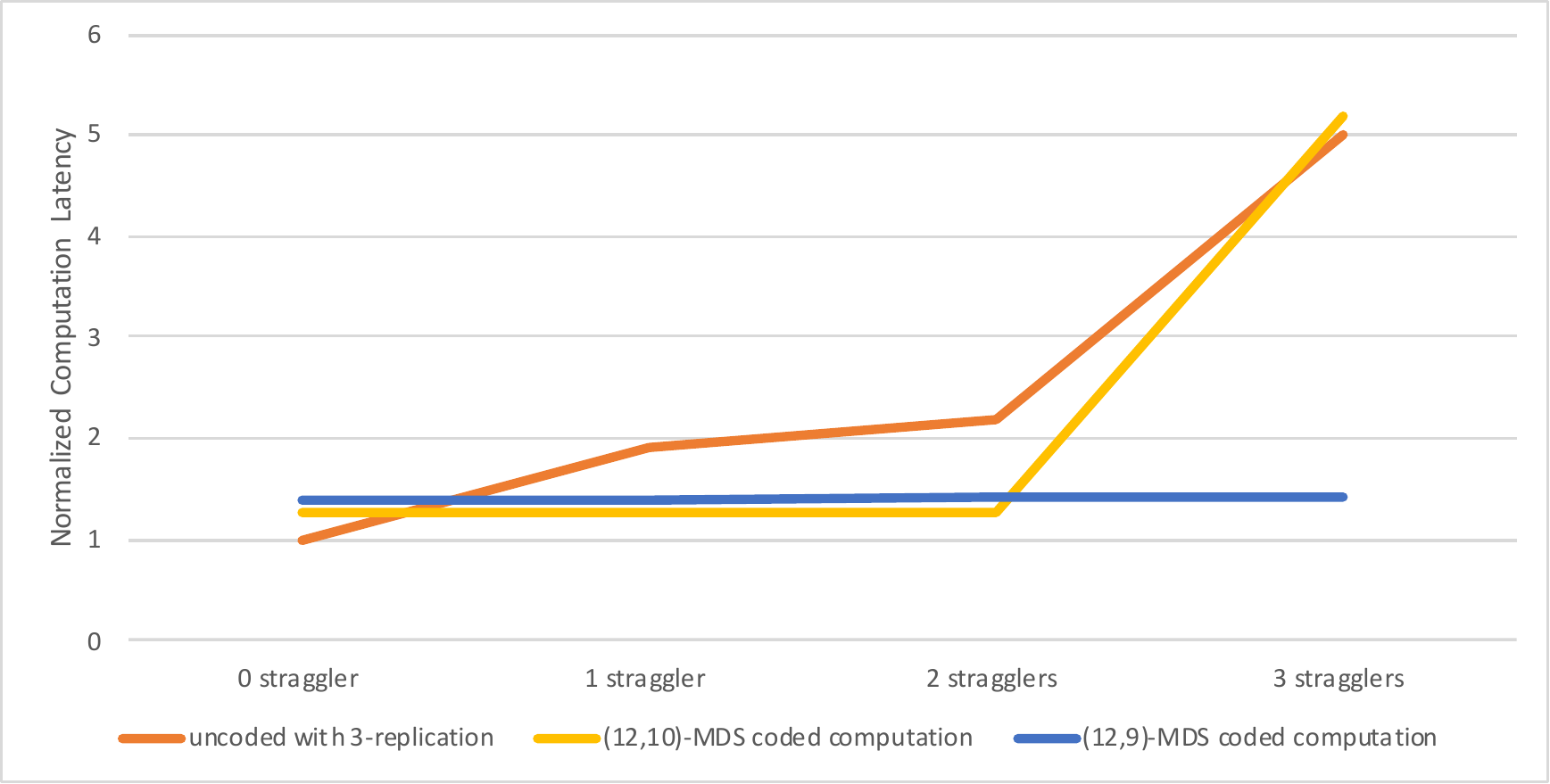}
    \caption{Logistic regression experiments}
    \label{fig:motivation}
\end{figure}

Figure \ref{fig:motivation} shows the computation latency in a cluster of 12 nodes with three schemes: uncoded with 3-replication, (12,10)-MDS coded computation, and (12,9)-MDS coded computation as the number of stragglers increases. In the uncoded with 3-replication strategy, if the number of straggler nodes is $r=3$ or more, computation latency increases significantly. The computation time of (12,10)-MDS coded computation increases exponentially when there are more than two stragglers. The computation time of (12,9)-MDS coded computation is constant with more stragglers. But there is an increase in baseline execution latency with this strategy compared to other schemes, because (12,9)-MDS code requires each worker node to perform more work than (12,10)-MDS code, even if the number of stragglers is fewer than 3.


In summary, although conservative MDS-coded computation can provide robust protections against stragglers, its computation overhead per node is higher and remains the same even when all the nodes in the cluster are fast, since it does not make efficient use of all worker nodes. These drawbacks bring us to our key idea which is to have a workload scheduling strategy that provides the same robustness as the $(n,k)$-MDS-coded computation, but only induces a much smaller computation overhead as if $(n, s)$-MDS-coded computation is being used when there are only $n-s$ stragglers in the cluster with $0 \leq (n-s) < (n-k)$. 

\begin{figure}
    \centering
    \includegraphics[width=\linewidth, height=3.5cm]{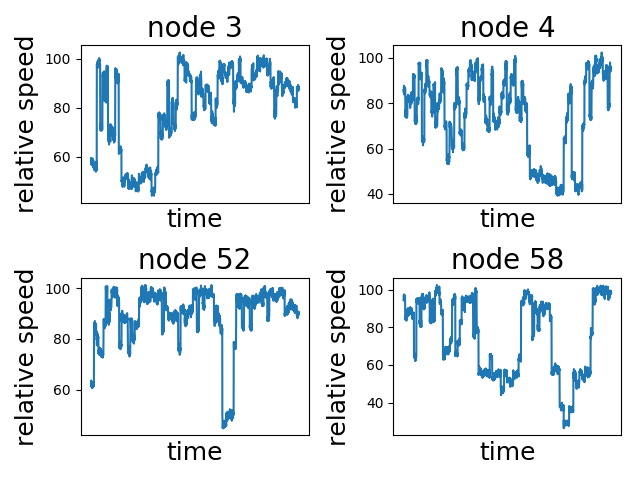}
    \caption{Measured speeds}
    \label{fig:measured}
\end{figure}

\subsection{Speed prediction and coded data} \label{speeds_Ref}
In the introduction section, we noted that it is important to consider the speed variations across compute nodes when determining the efficacy of discarding the work done by slow nodes in the MDS coded computing framework. To collect and analyze the execution speeds of servers, we conducted experiments on 100-compute nodes, referred to as droplets in Digital Ocean cloud \cite{digitalOcean}. Each droplet is similar to a t2.micro shared compute instance in Amazon AWS. For our experiments, each droplet node executes matrix-matrix multiplication and logs its execution times after completion of every 1\% of the task. The size of each matrix is 20000 by 5000. We analyzed the measured speeds at 1\% granularity intervals at all nodes. Figure \ref{fig:measured} shows the speed variations in 4 of the representative nodes. X-axis in each plot corresponds to time. Y-axis in each plot corresponds to speed of the node normalized by its maximum observed speed during the experiment.

One critical observation from the figure is that while the speed of each node varies over time, on average the speed observed at any time slot stays within 10\% for about 10 samples within the neighborhood. This relatively slow changing speed provides us an opportunity to estimate speeds of nodes in future intervals using speeds from past intervals. The speed estimates can be reasonably accurate for most of the time intervals except for a short time window when the speed changes drastically, but again we will soon be able to track the new speed as the nodes stay in that speed zone.

To find a good prediction mechanism, we considered the speeds of each node as a time series and modeled our problem as a time series forecasting problem. We evaluated several LSTM (Long Short-Term Memory \cite{lstm}) and Auto Regressive Integrated Moving Average (ARIMA) models to predict the speeds. The details of the speed prediction models are described in section \ref{subsec:LSTM}. The prediction accuracy of the LSTM model is better than the ARIMA models. As expected, only immediately after a large speed variance is observed the model prediction lags behind but catches up with the observed speed soon after.

\begin{figure}
    \centering
    \includegraphics[width=\linewidth, height=3.85cm, trim=4 4 4 4, clip]{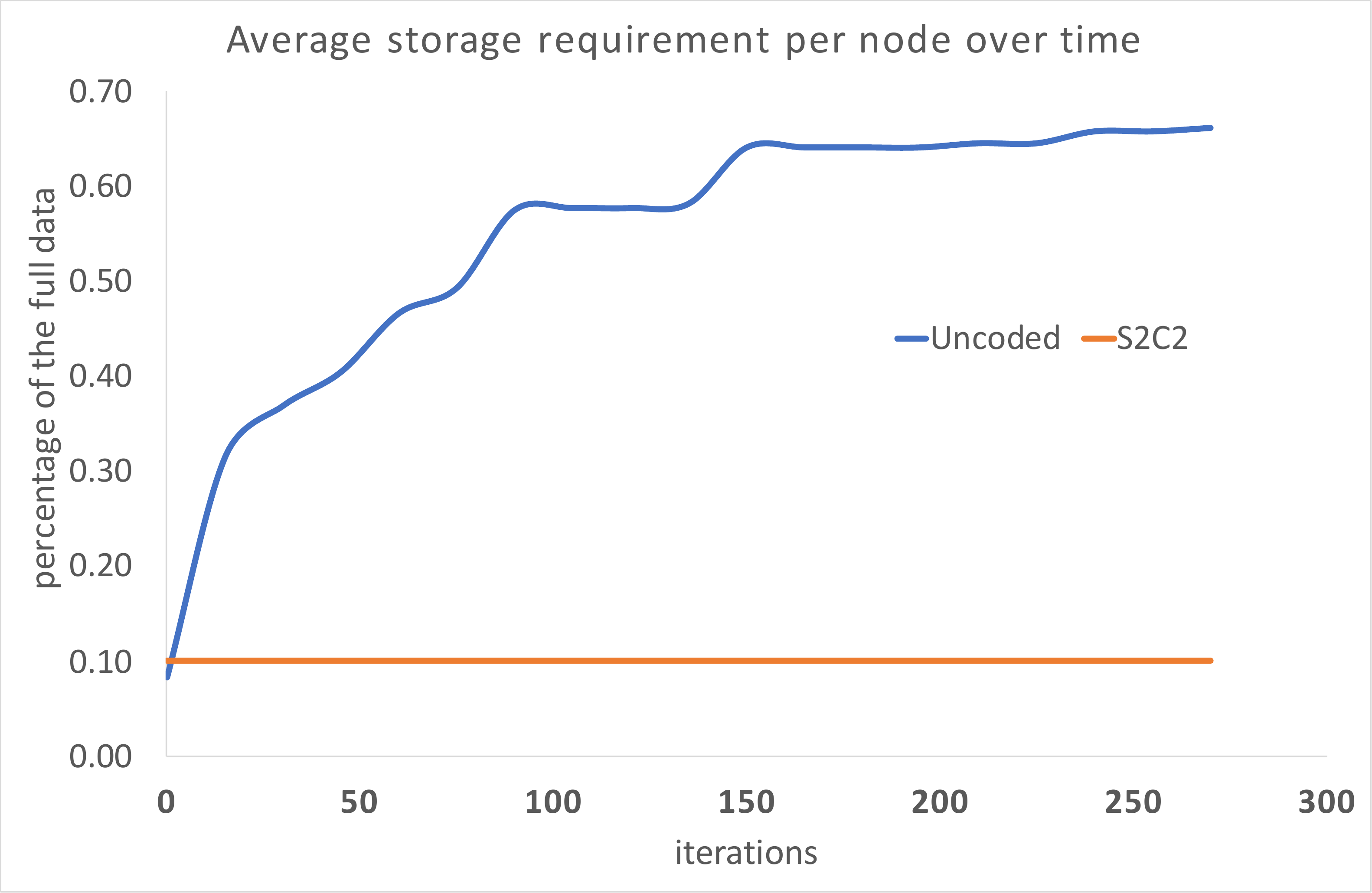}
    \caption{Storage overhead of uncoded computation even if we can predict speed for each node accurately}
    \label{fig:storage}
\end{figure}

Based on this critical observation we hypothesize that reliably estimating the speeds for next computation round allows master node to perform appropriate task assignment to the workers such that the computations performed by all workers can be utilized to obtain the final result. But this fine-granularity task assignment and utilization of all worker nodes becomes feasible only if there is no data movement overhead between rounds of computation. Coded computing is well suited for this fine-grained task assignment since the input data that is distributed among workers is encoded and as a result there would be no additional data movement needed between rounds of computation. However, this feature is not exploited in conventional MDS-coded computation. In uncoded computation, to assign workload optimally based on the predicted speeds, either each worker node will need to store significant percentage of the entire data, which can impose huge storage overhead; or it requires the master to redistribute the data among nodes at run-time, which can add huge communication overhead for iterative workloads such as gradient descent and page rank. To measure the storage overhead of uncoded computation, we performed experiments in our local cluster consisting of 12 worker nodes. We measure the total data moved to each node between rounds of computation and consider it as the effective storage needed at that node to avoid additional data movement. Figure \ref{fig:storage} shows the mean effective storage needed at each node to avoid data movement during the course of 270 gradient descent iterations for Logistic Regression. In this experiment, the uncoded computation has accurate predictions of speed of nodes for next iteration. It needs $67\%$ of the total data to be stored at each worker node to have zero data movement overhead. For $S^2C^2$ with (12,10)-MDS coding the data storage remains fixed at $10\%$ of the total data and much lower than the uncoded computation.


Following from these observations, we argue for $S^2C^2$ that exploits the unique feature of coded data availability and thereby utilize the compute capacity of all worker nodes. 

\section{$S^2C^2$}
\subsection{Basic $S^2C^2$ algorithm}
\begin{figure*}[h]
\centering
\begin{subfigure}{0.3\textwidth}
    \centering
    \includegraphics[trim ={1.3in 3.5in 1.45in 1.5in},     clip,width=.85\textwidth]{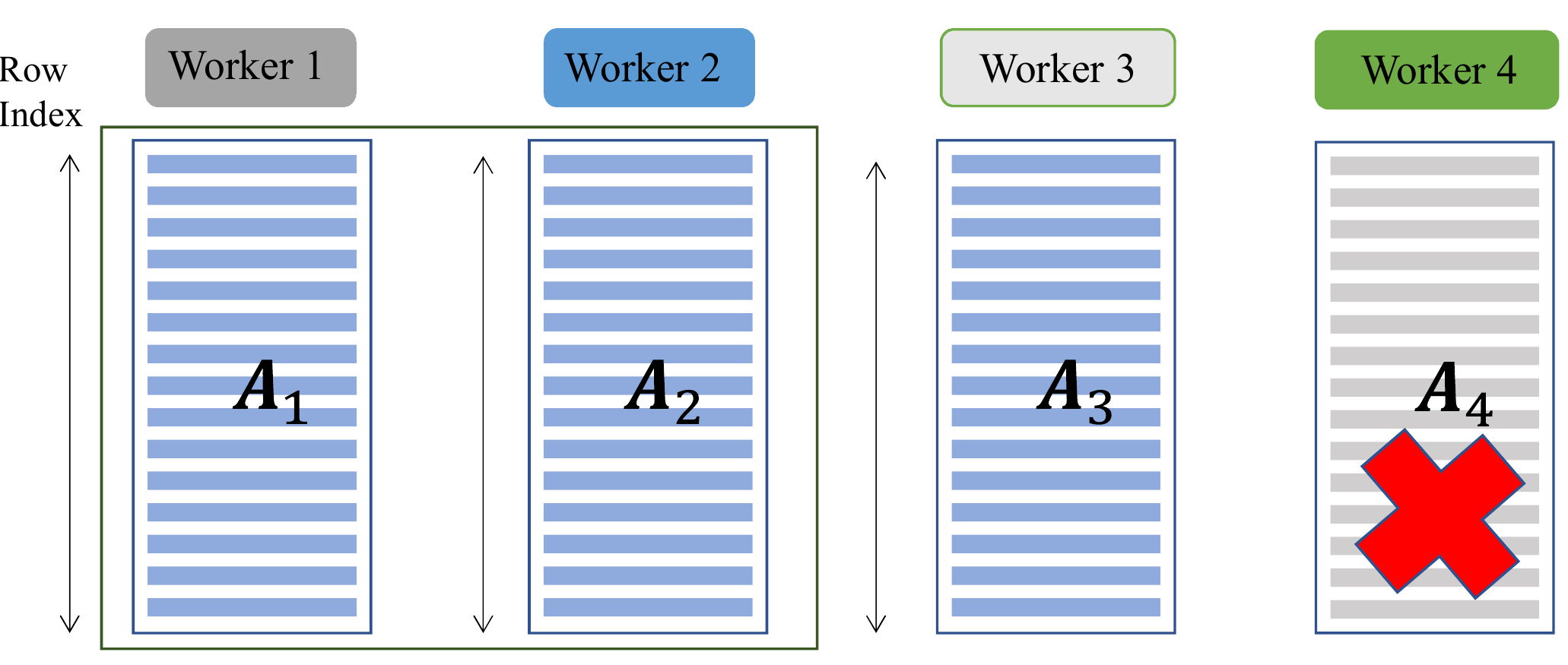}
    \caption{Conventional (4,2)-MDS.\\ Size of $A_1 = A_2 = A_3 = A_4 = \frac{A}{2}$ }
    \label{fig:regularMDS_4x2_1straggler}
\end{subfigure}%
\begin{subfigure}{0.3\textwidth}
    \centering
    \includegraphics[trim ={1.1in 3.5in 1.45in 1.5in},     clip,width=.85\textwidth]{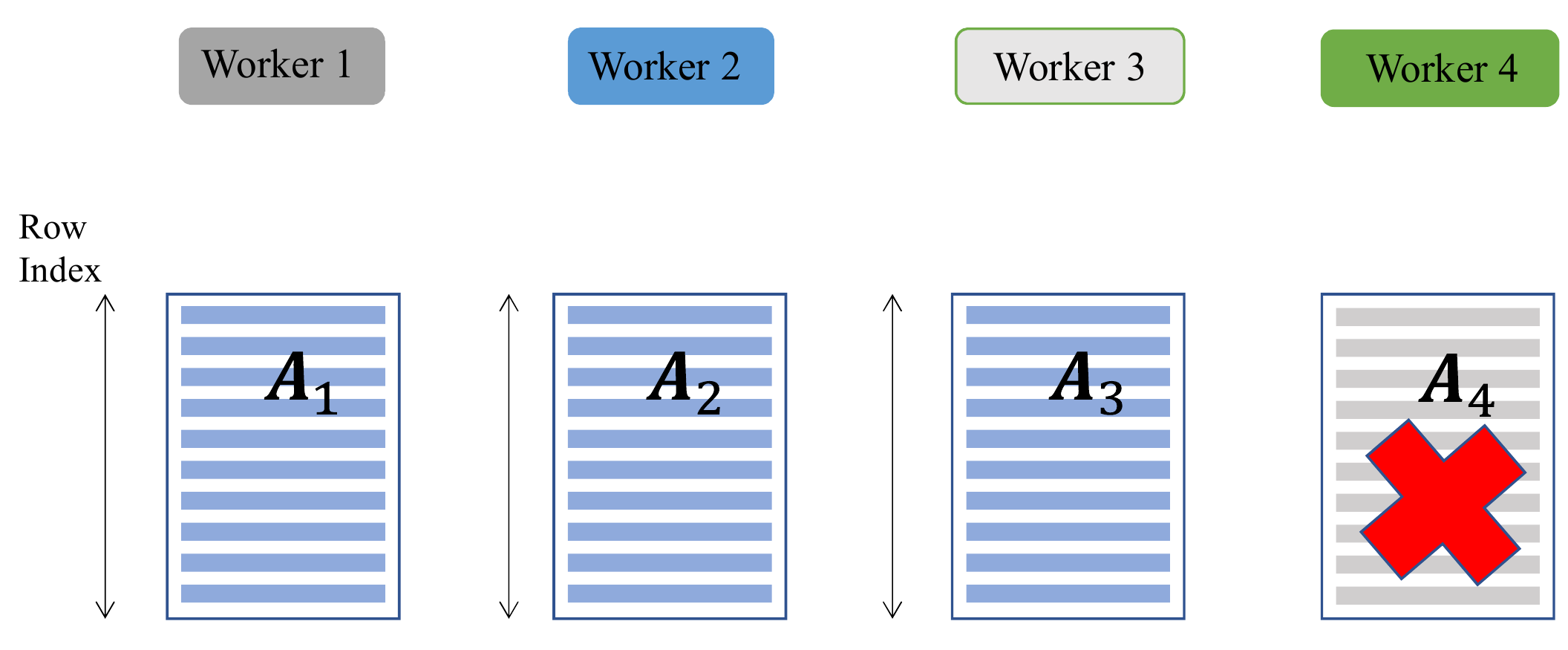}
    \caption{Conventional (4,3)-MDS.\\ Size of $A_1 = A_2 = A_3 = A_4 = \frac{A}{3}$}
    \label{fig:regularMDS_4x3_1straggler}
\end{subfigure}
\begin{subfigure}{0.3\textwidth}
    \centering
    \includegraphics[trim ={1.1in 3.5in 1.45in 1.5in},     clip,width=.9\textwidth]{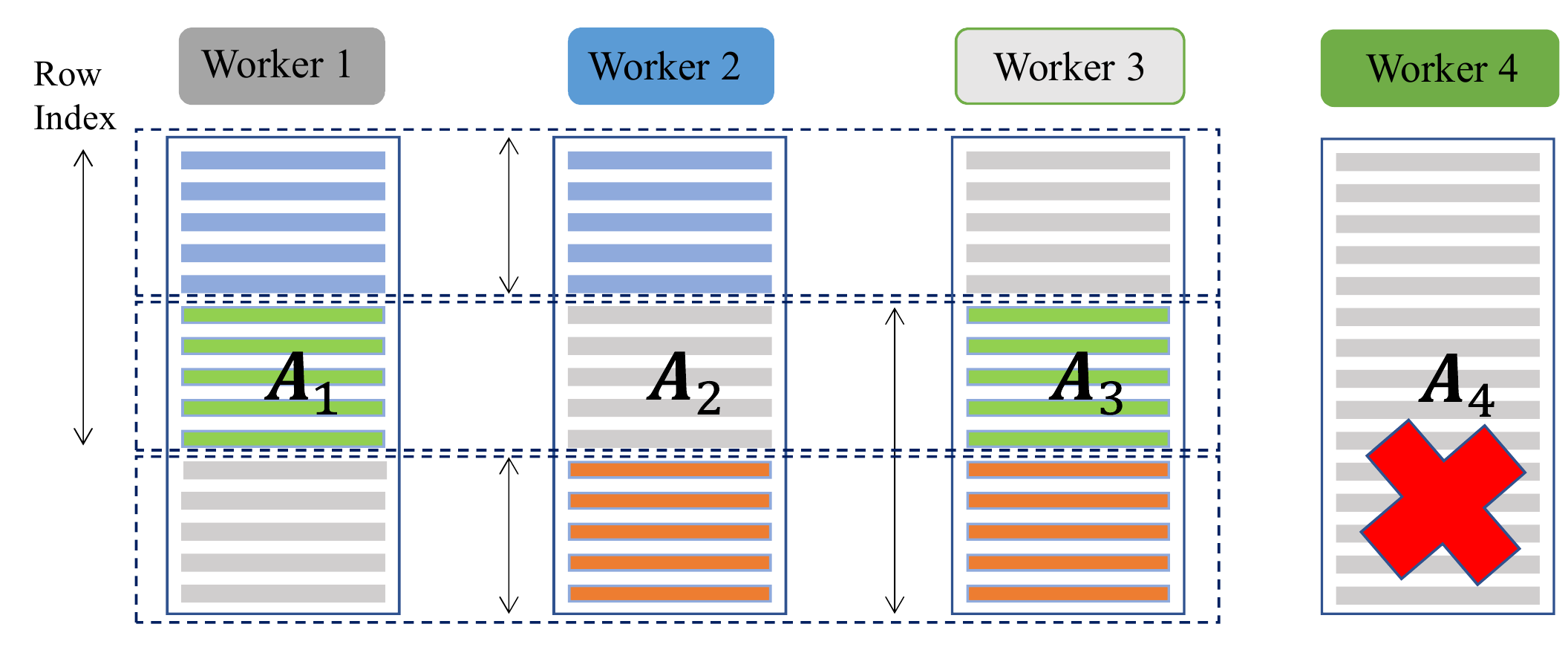}
    \caption{$S^2C^2$ on (4,2)-MDS.\\ Size of $A_1 = A_2 = A_3 = A_4 = \frac{A}{2}$. \\Work performed at each non-straggler worker = $\frac{A}{3}$}
    \label{fig:S2C2MDS_4x2_1straggler}
\end{subfigure}
\caption{$S^2C^2$ illustration on MDS codes}
\label{fig:$S^2C^2$}
\end{figure*}

\begin{figure*}[h]
\centering
\begin{subfigure}{0.5\textwidth}
    \centering
    \includegraphics[clip,width=.95\textwidth]{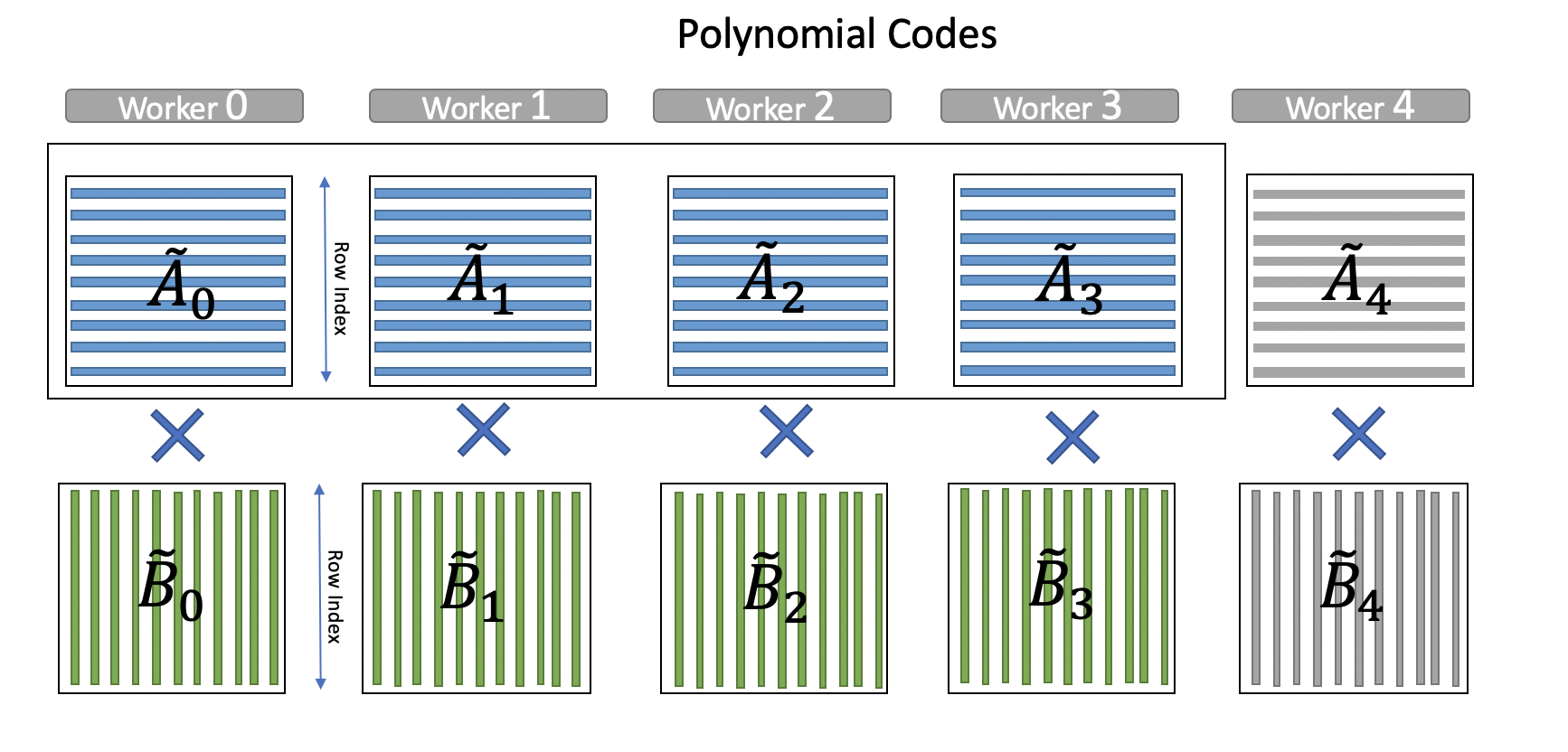}
    \label{fig:poly_regular}
\end{subfigure}%
\begin{subfigure}{0.5\textwidth}
    \centering
    \includegraphics[clip,width=.95\textwidth]{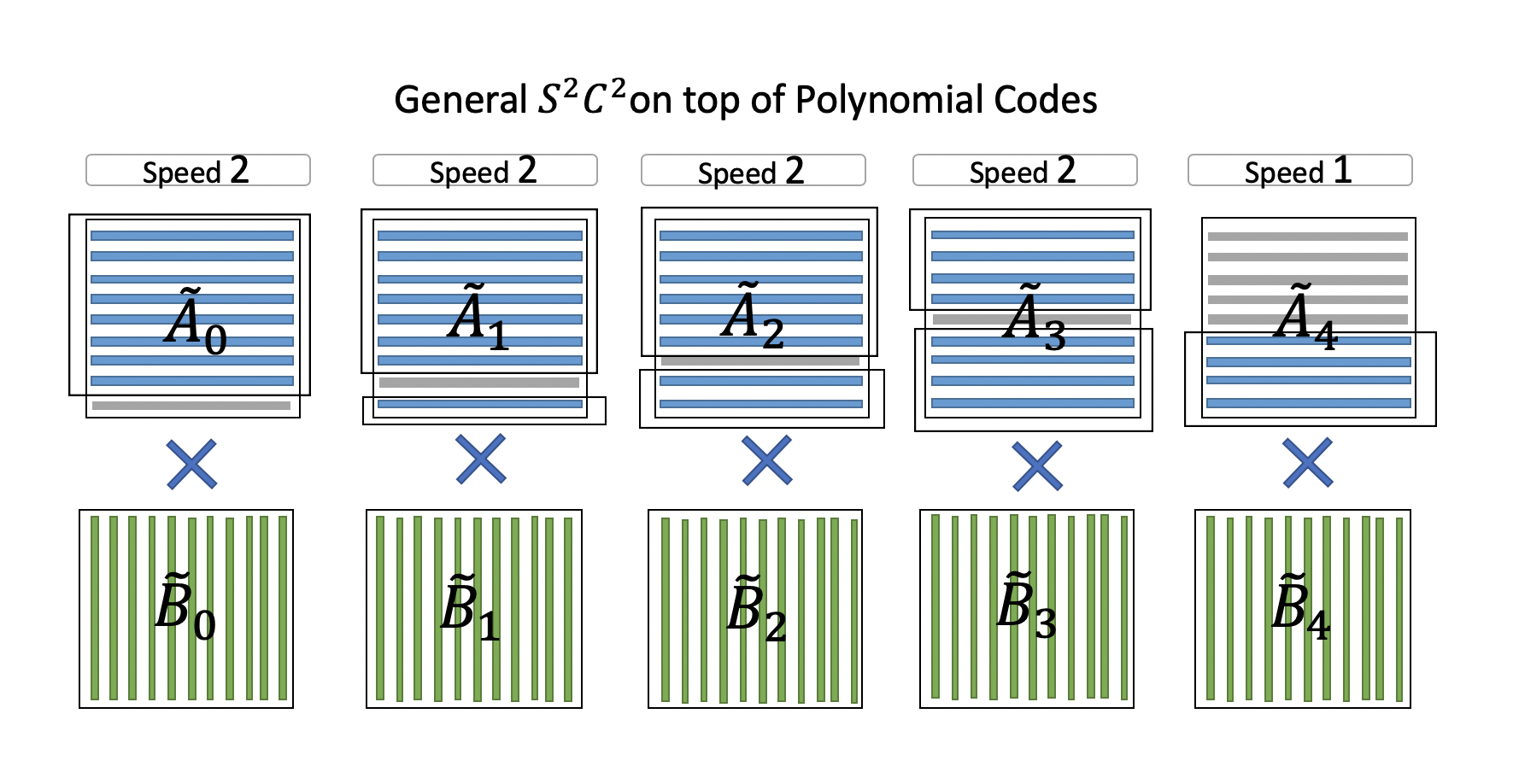}
    \label{fig:poly_s2c2}
\end{subfigure}
\caption{General $S^2C^2$ on polynomial codes}
\label{fig:poly_codes}
\end{figure*}
The major goals of the algorithm are to achieve high tolerance to stragglers and reduce computation work assigned per worker when the number of slow nodes observed during run time is less than the conservative estimate. To achieve high straggler tolerance the master node encodes and distributes the large matrix using a conservative $(n,k)$-MDS coding once at the beginning. To assign reduced computation work to worker nodes, master node then employs $S^2C^2$ algorithm. There are two key insights, in a cluster using conservative $(n,k)$-MDS coding, that underlie our algorithm.
\begin{itemize}
    \item Each worker node stores high redundancy encoded matrix data partition
    \item The master node can decode and construct the final product as long as it receives \textit{any} k out of n responses corresponding to each row index of partitioned matrix
\end{itemize}
Let there be \textit{n-s} $<$ \textit{n-k} stragglers in the (n,k)-MDS coded cluster. As we explained in previous section, when there are (n-s) stragglers (n,s)-MDS coding is the best suited coding strategy. But rather than using a new (n,s)-MDS code to re-encode the data, we use the  (n,k)-MDS coded data partition as is but we change the amount of work done by each node. In particular, $S^2C^2$ allocates \textit{decodable computational work assignment} per node equal to that in (n,s)-MDS coding instead of (n,k)-MDS coding. If $D$ is the number of rows in the original matrix, each node gets an allocation $\frac{k}{s} \times \frac{D}{k}$ = $\frac{D}{s}$ number of rows to be computed.

Figure \ref{fig:$S^2C^2$} provides an illustration of the $S^2C^2$ strategy in a cluster consisting of 4 worker nodes (and 1 master node). Figure \ref{fig:regularMDS_4x2_1straggler} shows the conventional (4,2)-MDS coded computation performed when worker 4 is the only straggler node and the remaining 3 workers have same speed. Note that (4,2)-MDS coding is conservative here, since it can support 2 stragglers but in this case there is only 1 straggler. Each worker node computes on its full partition but the master node needs only the results from workers 1 and 2 and can ignore the result from worker 3. Sub-matrices $\textbf{A}_1, \textbf{A}_2$ refer to the vertical divisions of the matrix $\textbf{A}$. Data stored in worker 3 is a coded matrix, $\textbf{A}_3 = \textbf{A}_1 + \textbf{A}_2$. Data stored in worker 4 is coded as $\textbf{A}_4 = \textbf{A}_1 + \textbf{2A}_2$. These codes are generated as per MDS-coding principles. 

In figure \ref{fig:regularMDS_4x3_1straggler}, conventional (4,3)-MDS coded computation when worker 4 is the straggler node is shown. Each non-straggler node computes on its full partition but the size of the partition here is smaller than partition size in the previous coding. The master node needs the results of all workers to construct the final product. Sub-matrices $\textbf{A}_1, \textbf{A}_2, \textbf{A}_3$ are the vertical divisions of the matrix $\textbf{A}$ into three parts. Data stored in worker 4 is coded as $\textbf{A}_4 = \textbf{A}_1 + \textbf{A}_2 + \textbf{A}_3$.

$S^2C^2$ with (4,2)-MDS Coded computation for this scenario is shown in figure \ref{fig:S2C2MDS_4x2_1straggler}. If we consider data in each worker as composed of 3 equal size partitions, worker node 1 computes only on the first and second of its partitions. Worker 2 computes only on the first and third of its partitions. Worker 3 computes only on the second and third of its partitions. As a result, each worker node is performing less amount of compute and this compute is equal to the amount performed by each worker in conventional (4,3)-MDS coded computation. Partitions to be computed at each worker are assigned to ensure that each row index is computed by exactly two workers. This is necessary for successfully decoding the results by the master node at the end of computation. For instance, worker node 3 computes on the middle third of matrix A3 (which is the coded A1+A2 matrix) and worker node 2 skips computing that portion of A2. As such the master has to decode the missing A2 from the computations performed by worker node 1 and worker node 3 to reconstruct the middle portion of A2. 

\subsection{General $S^2C^2$ Algorithm}
In cloud computing services and data centers, compute nodes within a cluster can have different speeds during run time, as described in section \ref{speeds_Ref}, due to them being shared or due to various micro-architectural events such as cache misses and other control/data bottlenecks. They can also be heterogeneous. We present a \textit{General} $S^2C^2$ algorithm which, unlike basic $S^2C^2$, can consider the variation in speeds of all nodes and assign work to them. 
At the beginning of execution of every application, matrix data is partitioned, encoded and distributed to the worker nodes using $(n,k)$-MDS coding. For efficient decoding and work allocation, general $S^2C^2$ algorithm also decomposes and considers each matrix partition as composed of chunks (groups) of rows i.e., over-decomposition. The speed predictions from LSTM model are provided to the general $S^2C^2$. Then workers are sorted according to their speeds. Starting from the worker with highest speed, each worker is assigned chunks to be computed equal to the ratio of its speed to the total available computational speed of all workers. If the assigned chunks for a worker turn out to be more than the total chunks in the partition already stored in a worker, the algorithm re-assigns these extra chunks to next worker. This case occurs when one worker is relatively much faster than all other workers. The algorithm is summarized in Algorithm \ref{alg:gen_S2C2}. It can be noted that general $S^2C^2$ algorithm uses relative speed predictions of the nodes during work allocation. In the scenarios where all non-straggler nodes have equal speed general $S^2C^2$ would reduce to basic $S^2C^2$. 

\subsection{Dealing with mis-prediction or failures}\label{misses}
Speed prediction algorithm can mis-predict when there is a sudden and significant change in the speeds of workers. Also, one of the worker nodes can die or fail during execution. To handle these scenarios, $S^2C^2$ algorithm employs a timeout mechanism. $S^2C^2$ collects results from the first $k$ workers that complete their work and measures their average response time. If the remaining $n-k$ workers do not respond within 15\% of the average response time, $S^2C^2$ considers this situation as a mis-prediction and reassigns the pending work among the $k$ completed workers. We choose 15\% based on the average error from our speed prediction algorithm (16.7\%). 

\subsection{Robustness of $S^2C^2$}
Coded computing with $S^2C^2$ is robust and can tolerate the same number of stragglers as conventional coded computing because:
\begin{itemize}
    \item Data distribution in $S^2C^2$ is identical to the data distribution in conventional coded computing.
    \item The worst case scenario may occur when the speed predictions for $S^2C^2$ completely fails. In this case the general $S^2C^2$ along with the timeout mechanism, described in section \ref{misses}, essentially turns into a conventional coded computing.
\end{itemize}

\section{Extension to bi-linear coded computing} \label{sec:poly_codes}
$S^2C^2$, being a workload distribution strategy, can be extended to many different coded computations. In this section we demonstrate how to apply it on top of the popular polynomial codes \cite{polyCodes}. We refer the reader to the paper for the mathematical underpinning of polynomial codes and we will only provide  brief overview to demonstrate its working and how $S^2C^2$ can be applied to such a generalized codes. The idea of polynomial codes is to encode data by computing polynomial functions. 

Consider computing  $\textbf{A}\textbf{B}$ on two matrices \textbf{A} and \textbf{B}, in a distributed manner using a cluster with $n$ nodes. Matrix $\textbf{A}$ is divided into $a$ sub-partitions along rows, and matrix $\textbf{B}$ is divided into $b$ sub-partitions along columns. Then $n$ encoded partitions each for $\textbf{A}, \textbf{B}$ are computed from these sub-partitions. Let us consider the scenario where $n=5$ nodes. In this scenario, $a=b=2$ i.e., each matrix has 2 sub-partitions. $A_0, A_1$ are sub-partitions of \textbf{A}. $B_0, B_1$ are sub-partitions of \textbf{B}. Computing $\textbf{A}\textbf{B}$ is composed of four partial computations $A_0 B_0, A_0 B_1, A_1 B_0, A_1 B_1$. Each encoded partition of \textbf{A} is of the form $\tilde{A_i} = A_0 + i A_1$ and each encoded partition of \textbf{B} is of the form $\tilde{B_i} = B_0 + i^2 B_1$, where $i$ is the node index $\in$ \{0,1,..n-1\}. In this scenario, node 0 stores $\tilde{A_0} = A_0 + 0. A_1$, $\tilde{B_0} = B_0 + 0. B_1$ and node 2 stores $\tilde{A_2} = A_0 + 2 A_1$, $\tilde{B_2} = B_0 + 2^2 B_1$, and so on. Each node computes product of it's two stored partitions. For instance, node 2 computes $A_0B_0 + 2 A_1B_0 + 2^2 A_0B_1 + 2^3 A_1B_1$. To fully decode the four partial computations, $A_0 B_0, A_0 B_1, A_1 B_0, A_1 B_1$, we need to get coded computation results from any 4 of the nodes. If none among the 5 nodes is a straggler, there is wastage of one node's computation similar to MDS coding. 


In figure \ref{fig:poly_codes} we illustrate how our $S^2C^2$ framework can be applied on top of such a polynomial coded bilinear computation. In figure~\ref{fig:poly_codes}, the cluster has $n=5$ nodes. For illustration purposes each matrix partition $\tilde{A_i}$ has 9 rows. A minimum of 4 responses per each row are needed for successful computation of $AB$. The relative speeds of each nodes are \{2,2,2,2,1\}. Node 4 is a partial-straggler. Conventional polynomial coded computing ignore the computation from this node. However, general $S^2C^2$ does not and it allocates partial work to it. General $S^2C^2$ allocates \{8,8,8,8,4\} rows to the 5 nodes respectively as highlighted by the bounding rectangles in each worker node. The last worker (speed 1) is shown to compute the last set of rows. Product of each row with $\tilde{B_i}$ is computed by exactly 4 nodes and sent to the master node.



\begin{algorithm}[h]
\caption{General $S^2C^2$ algorithm}
\label{alg:gen_S2C2}
\begin{algorithmic}
\State Lines with \# are comments
\State \textbf{Input:} List ($U$) of Speeds ($u_i$) of worker nodes, $n, k$ of coding, Number of rows per Node ($numRowsPerNode$)
\State \textbf{Output:} Computation assignment per node i $alloc_i$
\State \#over decompose each partition into chunks of rows 
\State $maxChunksPerNode = \sum u_i$
\State \#minimum total chunks needed for correct decoding 
\State $totalChunks = k \times maxChunksPerNode$
\State \#Sort the workers as per their speed $U$ in descending \State \#order and assign number of chunks to be computed
\For{each node $i$ in sorted $U$}
       \State \#Allocate number of chunks to node $i$ proportional to its speed 
       \State $chunksForNode[i]$ = ($\frac{u_i}{\sum\limits_{j=i}^{n} u_j} \times totalChunks$)
       \State \#Update total chunks left to be computed 
       \State $totalChunks$$=$$totalChunks-chunksForNode[i]$
\EndFor
\State \#Assign the exact chunks that will be computed
\State $chunkBegin=0$
\For{each node $i$ in sorted $U$}
       \State $chunkEnd = chunkBegin+chunksForNode[i]$
       \State $chunks\_node_i  \leftarrow [chunkBegin, chunkEnd]$
       \State $chunkBegin = chunkEnd\%maxChunksPerNode$
\EndFor
\State \#Convert chunks to exact row indices
\State $alloc_i =$ $convert(chunks\_node_i)$
\end{algorithmic}
\end{algorithm}

\section{Implementation}
At the beginning of computation, master node encodes the matrix data and distributes the encoded sub-matrices to the corresponding worker nodes. For MDS coding we are dealing with just a single matrix, but with Polynomial codes we have two matrices to encode, and both coding strategies use different encoding as described earlier. At the start of each iteration of our applications master node distributes the vector (${\overrightarrow x}$) to all worker nodes. At the end of each iteration, master node receives the sub-product from the worker nodes, decodes them and constructs the result vector.

Each worker node has two separate running processes, one for computation and one for communication. The computation process on the worker node performs the appropriate computation on encoded data, either a matrix-vector operation in MDS setting or a Hessian matrix computation in polynomial setting. The communication process is in charge of receiving input data from the master node, work assignment information, and sending the partial product, and controlling the start and stop of the computation process at the worker node.

\subsection{LSTM based speed prediction Model} \label{subsec:LSTM}
We used the speed data measured from our experiments in motivation section as the dataset for evaluating several prediction models. The train/test dataset split is 80:20. We evaluated several LSTM (Long Short-Term Memory \cite{lstm}) and Auto Regressive Integrated Moving Average (ARIMA) models to predict the speeds. Among ARIMA models, we evaluated using ARIMA(1,0,0), ARIMA(2,0,0) and ARIMA(1,1,1) models. We found that the ARIMA(1,0,0) model, which uses just the speed from past iteration, provided the highest prediction accuracy among all ARIMA models. Since this indicates that using the speed from the past iteration is enough, the evaluated LSTM models have 1 dimensional input. The dimension of hidden state is a hyper parameter and we experimented with different values. The best performing LSTM model consists of one single-layer LSTM with a hidden state being 4 dimensional with tanh activation, 1 dimensional input and output. The prediction accuracy of this LSTM model is better than the ARIMA(1,0,0) model. The LSTM model predicts the speeds of the nodes within $83.3$\% of their actual values. In statistical terms, the  Mean  Absolute  Percentage error of the model on the test set is $16.7$\%. This prediction error is better than ARIMA(1,0,0) by 5\%. This LSTM model is used to predict speed of nodes once every iteration. Input to the model is the speed of node from previous iteration and its output is the speed prediction for the next iteration. The LSTM model computation takes 200 microseconds for each node.

\subsection{$S^2C^2$ specifics}
Basic $S^2C^2$ strategy needs information on which nodes are stragglers. General $S^2C^2$ strategy needs information on the relative execution speeds of all nodes and it adjusts the work assignment to the worker nodes according to their speed. To obtain this information we rely on the iterative nature of our algorithms. Initially master node starts with the assumption that all the worker nodes have the same speed and this is provided as input to the current $S^2C^2$ strategy. The master then distributes the work assignment calculated by $S^2C^2$ to each worker node. Upon receiving the partial products from the worker nodes, master node also records the response time $t_i(iter)$ for each worker node $i$ corresponding to iteration $iter$. If the number of rows computed at worker $i$ is $\ell_i(iter)$, then the speed $s_i(iter)$ of each worker node for the current iteration is computed as $\frac{\ell_i(iter)}{t_i (iter)}$. These values from all nodes are provided as a batch input to the trained LSTM model which predicts speeds for the next iteration. The predicted speeds are fed into the General $S^2C^2$ strategy to generate the computational work assignment at each worker node for iteration ($iter+1$). Thus $S^2C^2$ automatically adapts to speed changes at the granularity of an iteration.

\subsection{Computing Applications}\label{subsec:LR_Ref}
We evaluated $S^2C^2$ on MDS using the following linear algebraic algorithms:  Logistic Regression, Support Vector Machine, Page Rank and Graph Filtering. Graph ranking algorithms like Page Rank and Graph signal processing algorithms employ repeated matrix-vector multiplication. Calculating page rank involves computing the eigenvector corresponding to the largest eigenvalue which is done using power iteration algorithm; Graph filtering operations such as the $n$-hop fitering operations employ $n$ iterations of matrix-vector multiplication over the combinatorial Laplacian matrix. We evaluate $S^2C^2$ on both these algorithms. We further evaluate $S^2C^2$ on polynomial coding for computing the Hessian matrix as described in ~\cite{oversketch}. The Hessian computation is of the form $\textbf{A}^T diagonal(x) \textbf{A}$, where $diagonal(x)$ refers to a matrix composed of elements of vector $x$ on its diagonal.


\subsection{System Setup}
We evaluated the above computing applications in a datacenter scale cloud setting in the Digital ocean cloud. On Digital ocean cloud we employ 11 shared compute instances each with 1 virtual CPU and 2 GB of memory. We use Kubernetes to bootstrap a cluster using these 11 nodes, with one being the master and the other 10 nodes being the worker nodes. We then dockerize the computing applications and deploy them on the cloud.

\subsection{Verification in a controlled cluster}
For theoretical verification purposes we also evaluated all the applications and results on our local cluster where we had the ability to precisely control the straggler behavior. Our local cluster is composed of 13 identical servers. Each server consists of two Intel Xeon CPU E5-2630 v3 each with 8 cores (8 threads, 2.40 GHz),  20 MB of L3 cache, running Centos linux version 7.2.1511. Each machine has 64GB of DRAM. All the servers have access to a shared storage node of size 1 TB. All the servers are connected to one another through Mellanox SX6036 FDR14 InfiniBand switch with a bandwidth of 56 Gbps. We use one of the nodes as master node and other 12 nodes as worker nodes.

\section{Evaluation}\label{evaluation}
\subsection{Results from controlled cluster} \label{localcluster}

\textbf{Baseline strategies:}
We implemented and evaluated two baseline strategies in our controlled cluster environment: Our first baseline is an enhanced Hadoop-like uncoded approach that is similar to LATE~\cite{late}. In this baseline we used a 3-repetition strategy with up to six tasks that are speculatively launched. The strategy provides 3 copies of data at 3 randomly selected nodes in the distributed system. This enhanced Hadoop strategy does not enforce strict data locality during speculation, unlike traditional Hadoop, and allows data to be moved at runtime if a task needs to relaunched on a node that does not have a copy of the data. We allow up to six tasks to be speculatively launched. Furthermore, the speculative task  assignment strategy always tries to find a node that already has a copy of the data before moving the data, thereby allowing data communication only when absolutely needed.

The second baseline is the MDS-coded computation proposed in prior work~\cite{speedUpML} and described previously in section \ref{background}. The two MDS-coding schemes we evaluated in the controlled cluster are: $(12,6)$-MDS as the conservative scheme, and $(12,10)$-MDS as the optimistic scheme. No data movement is allowed in these schemes during computation. The purpose of showing results for $(12,6)$-MDS coding is simply to show the robustness of our scheme in the presence of such high redundancy. We do not expect that system designers will provision 2x computation redundancy in practice. Hence, we will highlight $(12,10)$-MDS results in our discussion in the next section.

\begin{figure}
    \centering
    \includegraphics[width=\linewidth, height=4.5cm, trim=4 4 4 4, clip]{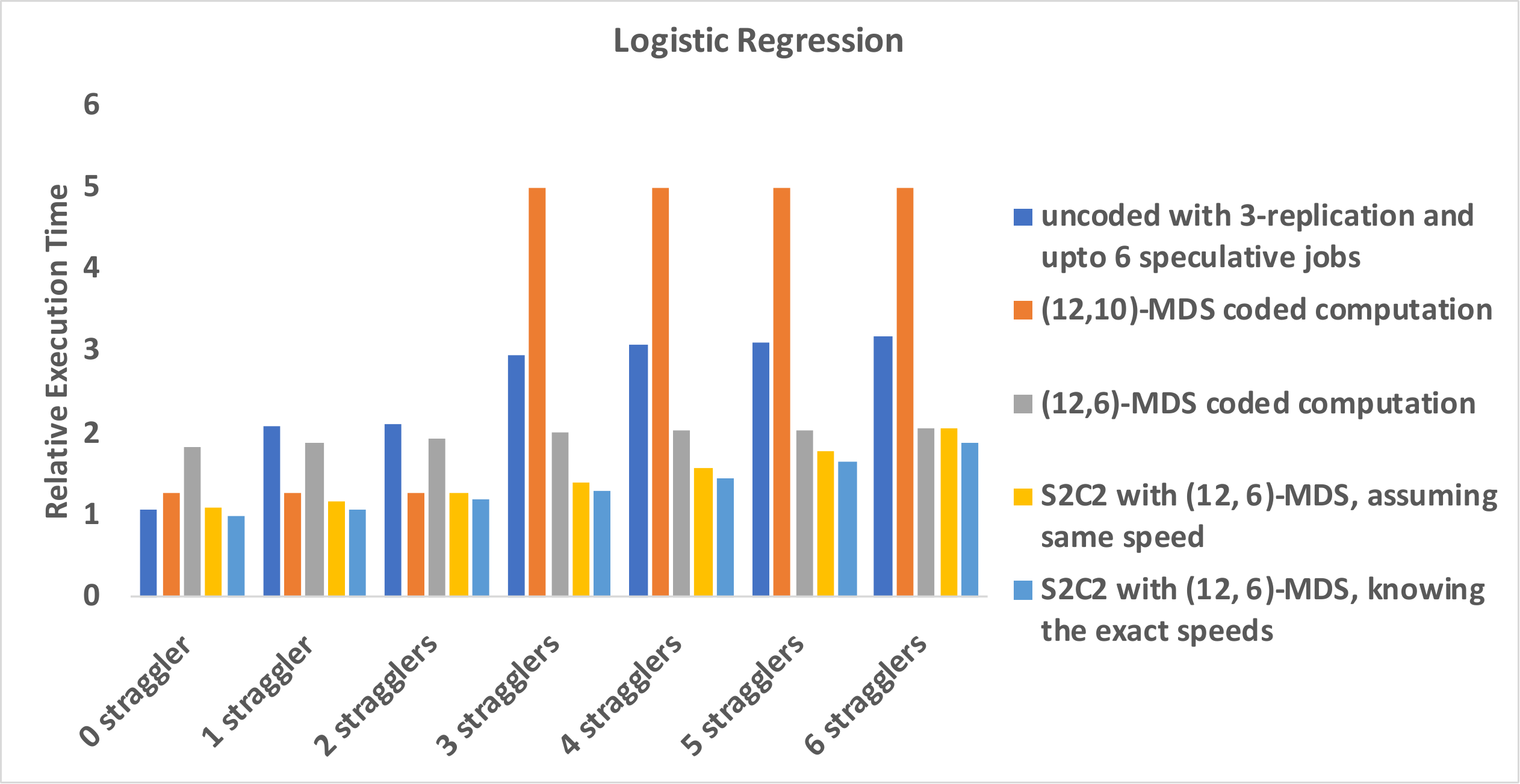}
    \caption{LR execution time comparison} 
    \label{fig:LR_vary}
\end{figure}

\begin{figure}
    \centering
    \includegraphics[width=\linewidth, height=4.5cm, trim=4 4 4 4, clip]{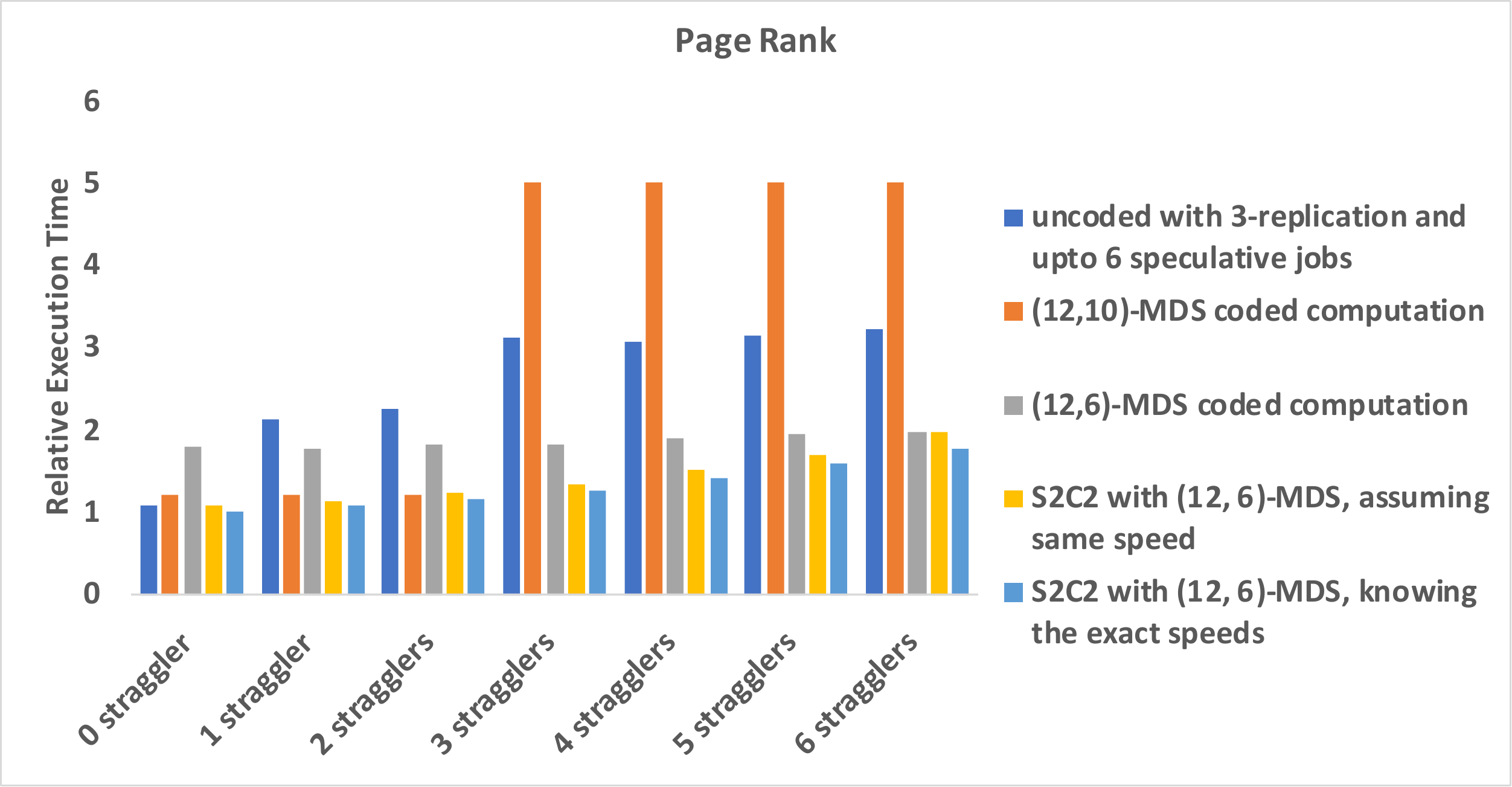}
    \caption{PR execution time comparison}
    \label{fig:PR_vary}
\end{figure}

\noindent \textbf{Results:}
We evaluated the performance of $S^2C^2$ against the baseline strategies for varying straggler counts in our 12-worker-node cluster and these different cases correspond to the X-axis in the figures \ref{fig:LR_vary} and \ref{fig:PR_vary}. Each bar in the plots captures the average relative execution time spent by the application for 15 iterations, normalized by the execution time of the uncoded strategy when there is 0 straggler in the cluster. The execution time includes the time worker nodes spend computing on their data partitions, the time spent in communication between master and worker nodes, time spent by the master node in decoding the results from workers. The encoding and distribution of matrix data are not shown in the figures as it is a one-time cost.

\subsubsection{Logistic Regression and SVM}
We evaluated gradient descent for logistic regression (LR) and SVM. The results for both of them are very similar and hence we focus the discussion on evaluations of LR. For our experiments we use publicly available gisette dataset from UC Irvine machine learning repository \cite{uci}. The data in this dataset is duplicated to create a larger dataset. The final size of data partition in each node is 760 MB. All the worker nodes pre-load the assigned data partitions into memory before beginning the computation. Only one processor thread in each worker node is used for computation.

In our controlled cluster environment, we define a straggler as a node that is at least 5x slower than the fastest performing node. And even non-straggler nodes may have up to $20\%$ variation between their processing speeds. We compare the three baselines with the two versions of $S^2C^2$: basic $S^2C^2$ that does not consider this $20\%$ variation in speeds of the non-straggler workers and treats all the non-straggler workers as having equal speed, and general $S^2C^2$ algorithm takes the $20\%$ speed variation into account and allocates different computational work to non-straggler workers accordingly. The results are shown in Figure \ref{fig:LR_vary}.

As shown in the figure \ref{fig:LR_vary}, when there are no stragglers, all strategies have low execution times with $S^2C^2$ having the lowest. The generalized $S^2C^2$ algorithm has the lowest execution time even with zero stragglers because it takes advantage of the 20\% speed variations to assign different amounts of work to different nodes.  
As the number of stragglers increases, the execution time of uncoded strategy increases since the slower job needs to be detected and re-executed. Whereas in coding based strategies there is no need for re-execution. Once the number of stragglers exceeds 2, the uncoded strategy's performance starts to degrade and it is 3x of the execution time compared to no straggler scenario. The super linear degradation is because data partition will need to be moved across worker nodes prior to the re-execution and communication costs start to play a role in the overall performance loss. Note that when the number of stragglers gets closer to the replication count then there is a higher probability that the node where the re-execution happens does not have a replica. Hence, data movement is in the critical path of execution.  

For the $(12,10)$-MDS coded computation, the execution time remains steady with one and two stragglers but grows super linearly once the straggler count exceeds two,  since it is designed to protect against a maximum of only two stragglers. A more redundant MDS coding strategy is the primary option to deal with higher number of stragglers. Hence, ideally the programmers should not be burdened with choosing an aggressive code with less redundancy and then have to pay significant penalty if the selected redundant code is not enough.  $S^2C^2$ solves this problem by allowing the programmers to select a more aggressive redundancy in their codes and yet not pay the penalty when there are fewer stragglers as shown in the $(12,6)$-MDS code result. 

Both versions of (12,6)-MDS coded $S^2C^2$  not only are able to provide robustness against up to two stragglers in the cluster, but also are able to reduce the computation overhead due to the use of coding when there are fewer or no stragglers in the cluster. By taking the various speeds of the non-straggler worker nodes into account, the general version of the $S^2C^2$ strategy is able to outperform the conservative (12,6)-MDS coded computation strategy even more than the basic version of $S^2C^2$. This result indicates that even if we can't take into account the precise variation in the processing speeds of various non-straggler nodes, the basic $S^2C^2$ algorithm provides excellent performance and robustness. However, if the processing speed information is more accurately gathered, the generalized $S^2C^2$ can squeeze the hidden compute slack in the 20\% speed variation and provide further performance improvements without compromising robustness.

\subsubsection{Page Rank and Graph Filtering}
We evaluated page rank (PR) and graph filtering. The results for both of them are very similar and hence we focus the discussion on Page Rank. We used the ranking dataset available from \cite{utoronto}. This dataset is duplicated to create a larger dataset that is used in evaluation.

The execution time for page rank is plotted in Figure \ref{fig:PR_vary}. Similar to logistic regression results,  $S^2C^2$ algorithms significantly outperform the baseline strategies. The general $S^2C^2$ algorithm reduces the execution time compared to basic $S^2C^2$ in all scenarios.

\subsection{Results from industrial cloud deployment} \label{real_deploy}

\begin{figure}
    \centering
    \includegraphics[width=\linewidth, height=4cm, trim=4 4 4 4, clip]{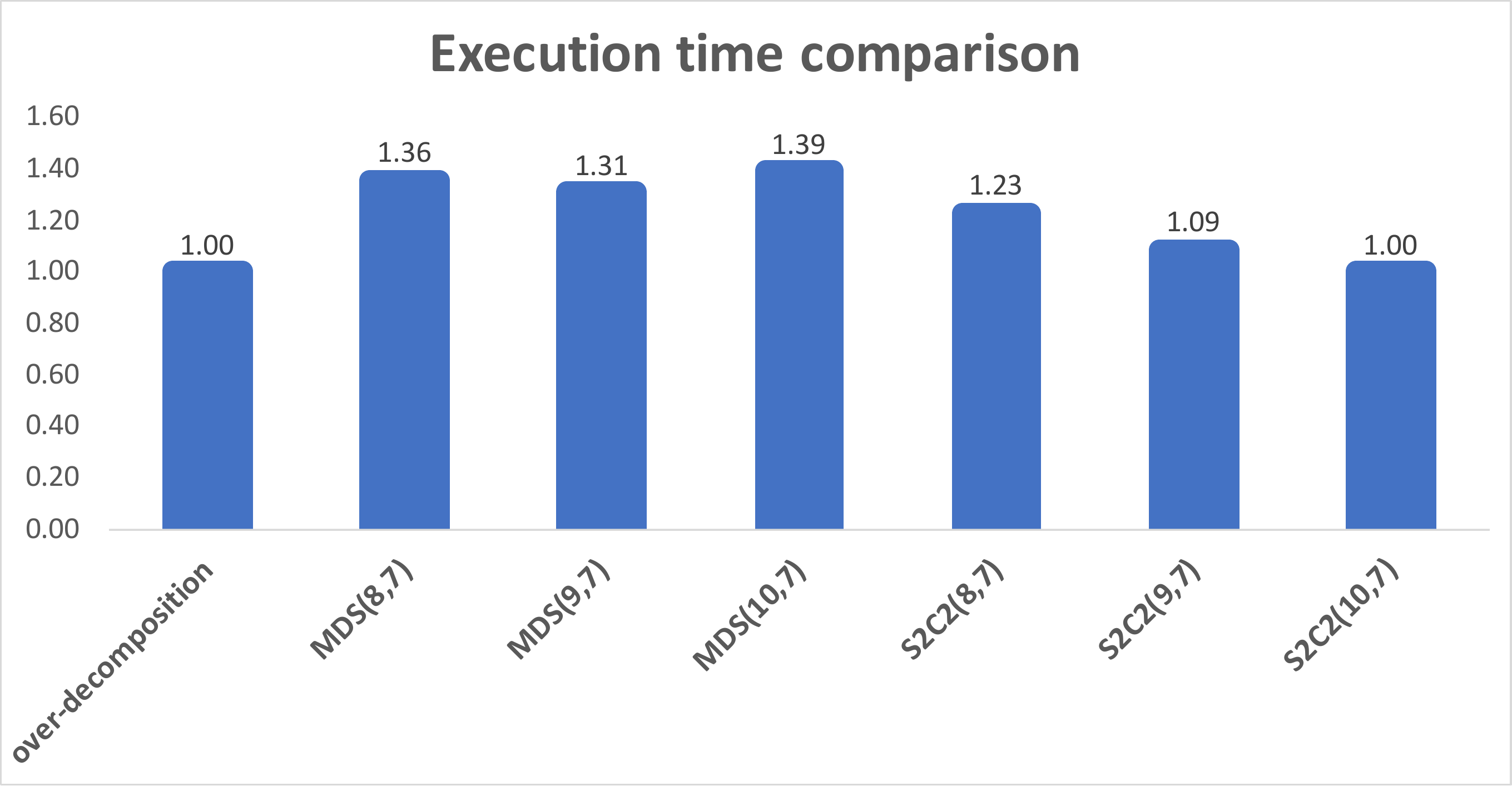}
    \caption{Execution time comparison on cloud when $S^2C^2$ has low mis-prediction rate}
    \label{fig:perfCloudBest}
\end{figure}

\begin{figure}
    \centering
    \includegraphics[width=\linewidth, height=3.5cm, trim=4 4 4 4, clip]{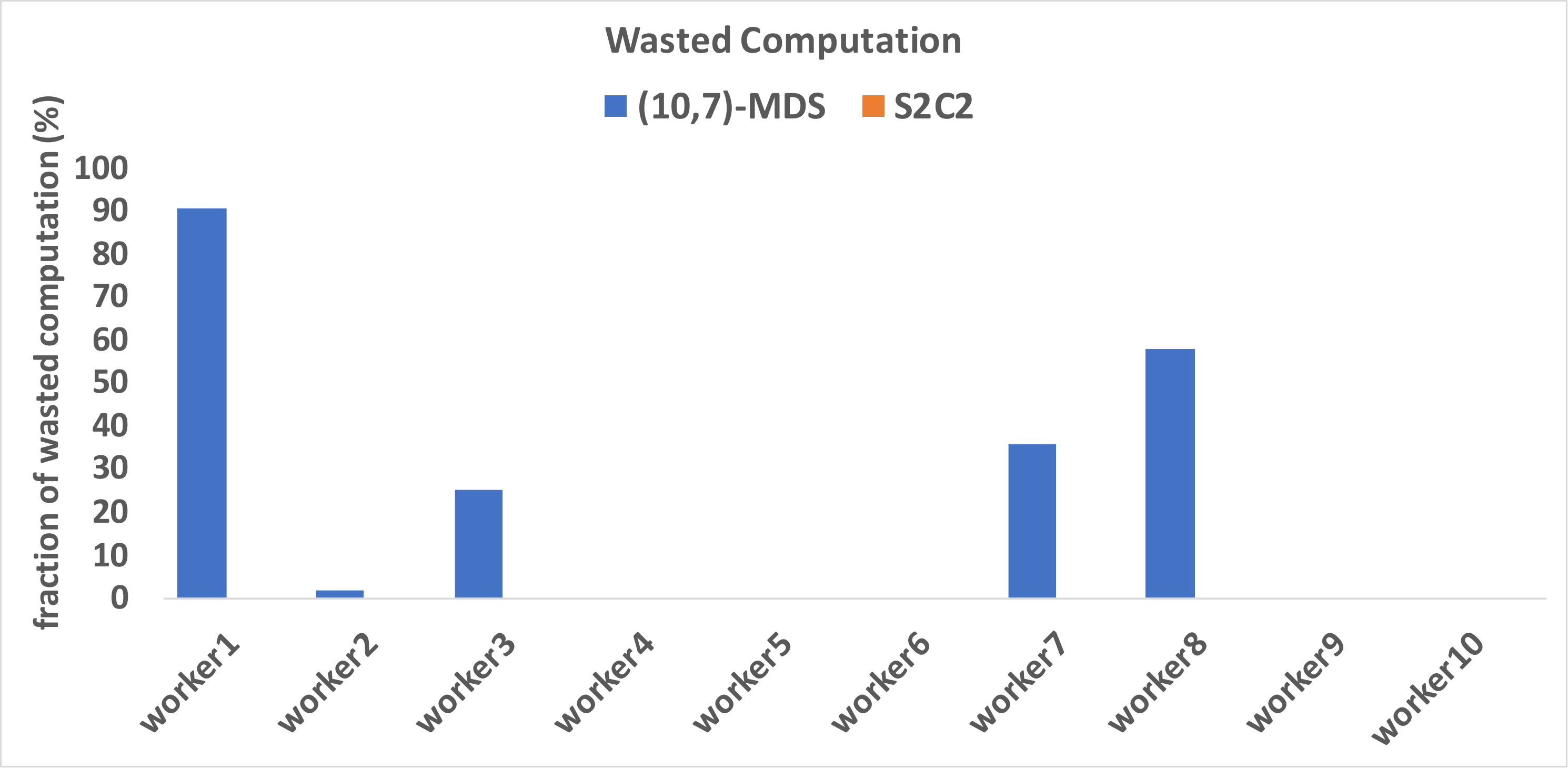}
    \caption{Per worker wasted computation effort with low mis-prediction rate}
    \label{fig:wastedWorkerBest}
\end{figure}

In this section we discuss results from our experiments on Digital Ocean cloud. Note that in this setup we no longer can  control the speed variations or the presence or absence of a straggler. Instead we simply rely on the inherent speed variations of the 10 worker nodes we used in the cloud environment to quantify the benefits of $S^2C^2$. In our experiments we evaluate and compare the performances of general $S^2C^2$ strategy against MDS coded computation and an over decomposition strategy based on Charm++~\cite{CharmppOOPSLA93,sc14charm}. We evaluated $S^2C^2$ and MDS coded computation under (10,7), (9,7) and (8,7)-MDS codes. 

\textbf{Charm++ based over-decomposition baseline:}
In the cloud setting, we evaluated an over-decomposition baseline strategy inspired by charm++~\cite{CharmppOOPSLA93,sc14charm}. In our implementation we combine over decomposition and speed prediction. We over-decompose each data partition by a factor of 4. The data is divided into $40$ partitions with each of the $10$ workers receiving 4 partitions. The data is also replicated by a factor of 1.42, similar to replication in (10,7)-MDS coding. The additional partitions are distributed in a round-robin fashion across the $10$ workers. Master node uses predictions from the speed model to do load balancing and transfer of partitions between workers during computations. This is better than the uncoded baseline strategy used in our controlled cluster environment since it allows for finer grained data transfer.

During the course of our experiments we observed different mis-prediction rates from the LSTM speed prediction model. We show and discuss the performance gains from the experimental conditions where we observe the best and worst case mis-prediction rates. The performance results obtained across various applications are similar (as has been shown also in the local cluster setting). We focus on the SVM results in this section.
\subsubsection{Results in low mis-prediction rate environment}
The average relative execution times for 15 iterations of SVM are shown in figure \ref{fig:perfCloudBest} when we observe a 0\% mis-prediction rate for worker speeds. Generally this happens when there are no significant variations in speeds between the nodes. The execution times of all strategies are normalized by the execution time of $(10,7)$-$S^2C^2$. First we can observe that over decomposition approach performs better than MDS coded computation. This result is expected since over-decomposition strategy utilizes all the 10 worker nodes to compute the result and each worker processes 1/10 of the data. But each worker in the (10,7), (9,7) and (8,7)  MDS-coded computation scenarios process 1/7th of the data. Next, we observe that all three variations of MDS-coded computation show similar execution times. In all cases the work performed by a single worker remains same and only the results from fastest 7 workers are used by the master. Over decomposition performs similar to (10,7)-$S^2C^2$ in this environment since there is no additional data movement during computations.

For all 3 data coding variations $S^2C^2$ outperforms regular MDS coded computation. Further, performance of $S^2C^2$ increases as the redundancy is increased. This is because work done in a single worker decreases as redundancy is increased. (10,7)-$S^2C^2$ outperforms the (10,7)-MDS coded computation by $39.3\%$. (10,7)-$S^2C^2$ performs best over (10,7)-MDS coded computation when all 10 workers are always fast during execution; in this scenario $S^2C^2$ uses all 10 worker nodes while MDS still relies only on 7 worker nodes. The exact reduction would be $\frac{10-7}{7}=42.8\%$. $S^2C^2$ with $0\%$ mis-prediction rate captures this best possible reduction in execution time.

Figure \ref{fig:wastedWorkerBest} plots the wasted computation measured in each of the worker node during execution of the conservative (10,7)-MDS coded computation and (10,7)-$S^2C^2$. Since the mis-prediction rate is 0\% there is no wasted computation effort in $S^2C^2$. In this execution, workers 1, 3, 7 and 8 have high wasted computation with (10,7)-MDS coded computation. Worker 1 has close to 90\% of its computation wasted. Further analysis showed that in this experiment worker 1 is only slightly slower than the fastest 7 workers but the MDS-coded computation ignores the execution of the 3 remaining workers after it receives results from the fastest 7 workers.

\begin{figure}
    \centering
    \includegraphics[width=\linewidth, height=4cm, trim=4 4 4 4, clip]{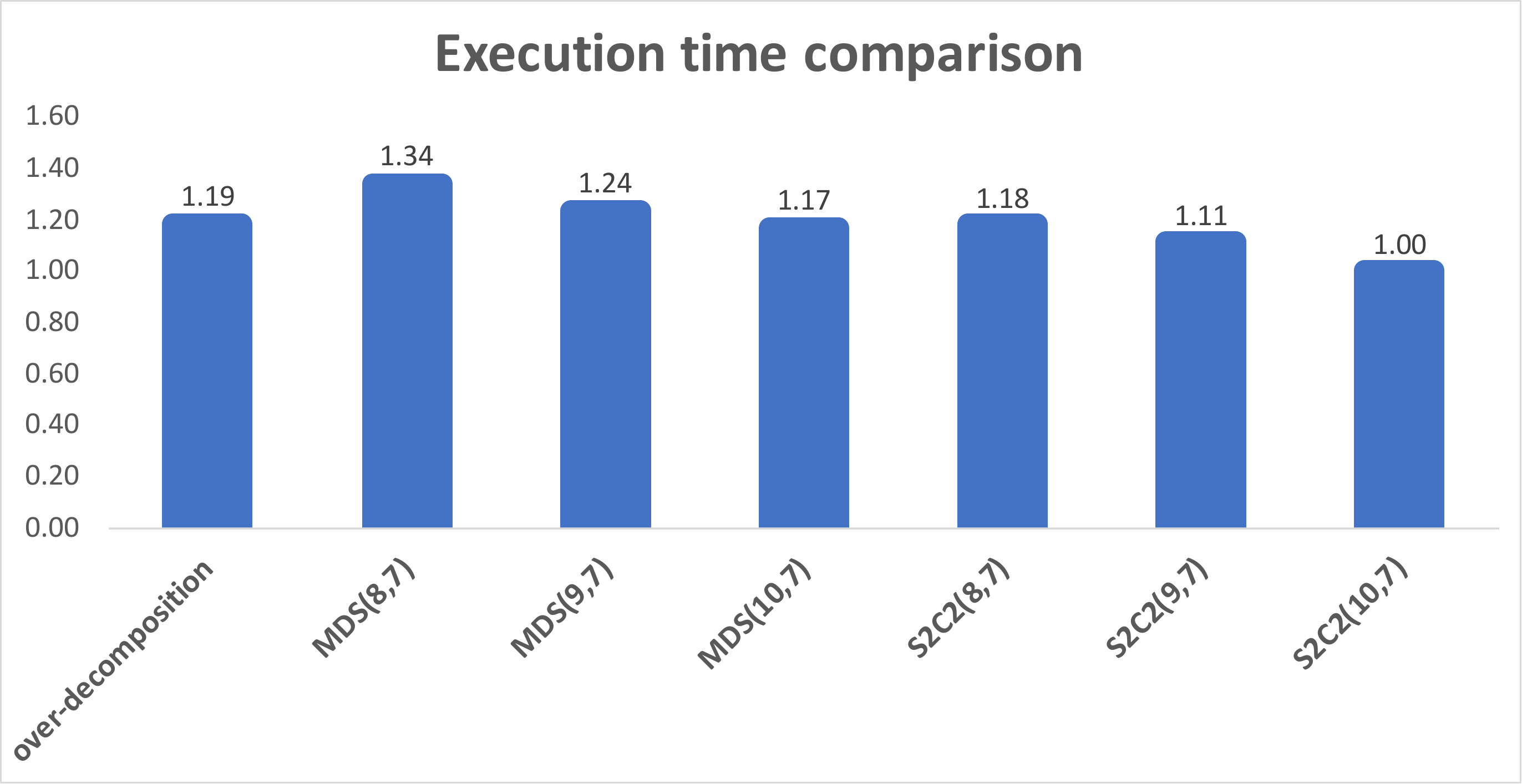}
    \caption{Execution time comparison on cloud when $S^2C^2$ has high mis-prediction rate}
    \label{fig:perfCloudWorst}
\end{figure}

\begin{figure}
    \centering
    \includegraphics[width=\linewidth, height=3.5cm, trim=4 4 4 4, clip]{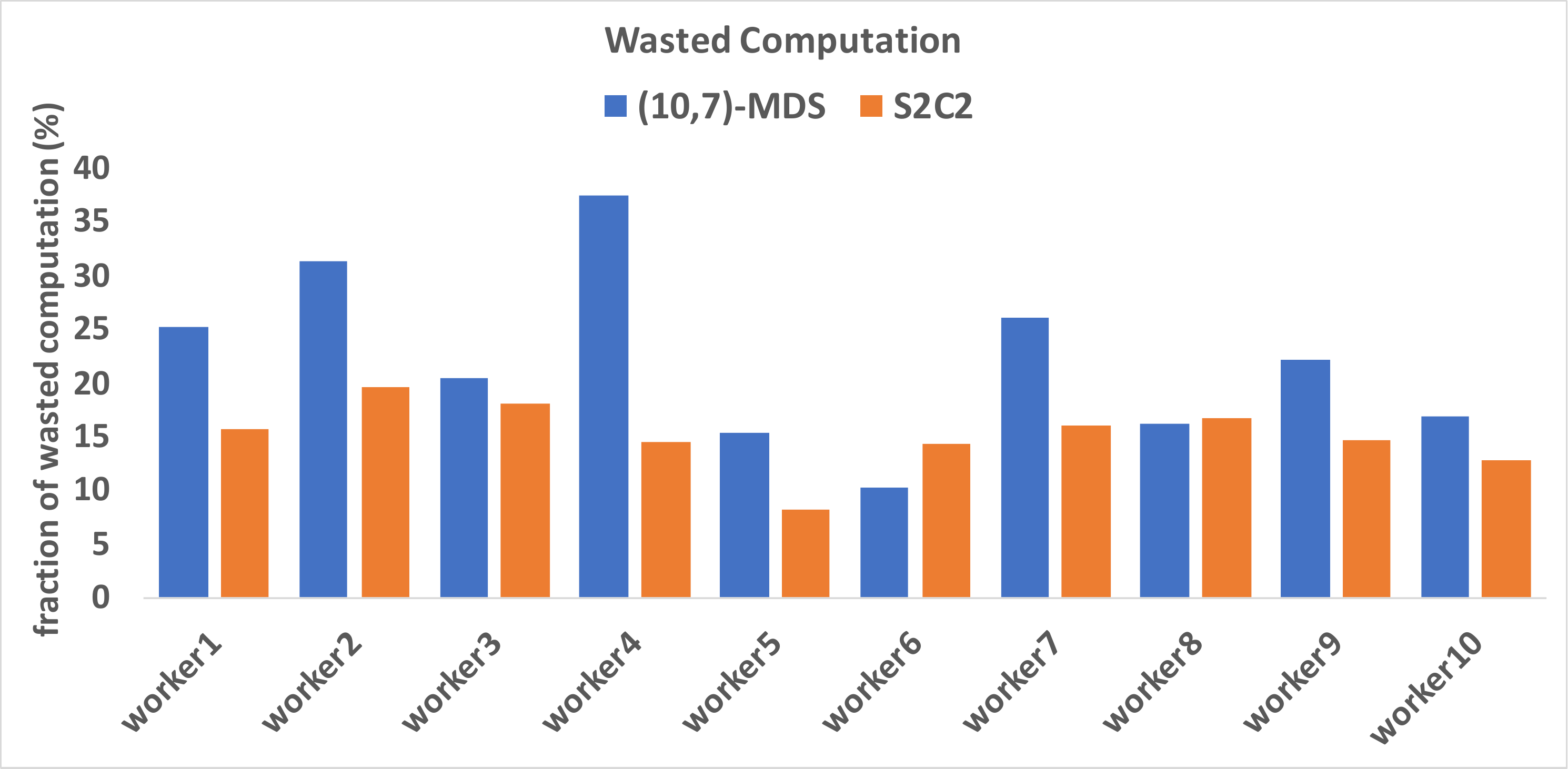}
    \caption{Per worker wasted computation effort with high mis-prediction rate}
    \label{fig:wastedWorkerWorst}
\end{figure}

\subsubsection{Results in high mis-prediction rate environment}
During our experiments with shared VM instances on DigitalOcean, we observe the highest mis-prediction rate is 18\%. Generally these mispredictions happens when there are significant and sudden variations in speeds over time. Under this condition, the average execution times for 15 iterations of SVM are shown in figure \ref{fig:perfCloudWorst}. (10,7)-MDS coded computation performs better than (9,7) and (8,7)-MDS coded computation because the probability of any 7 out of 10 nodes being fast is higher than any 7 out of 9 nodes or 7 out of 8 nodes being fast. (8,7)-$S^2C^2$ outperforms (8,7)-MDS coded computing by 13\%, whereas (9,7)-$S^2C^2$ outperforms (9,7)-MDS coded computing by 11\% and (10,7)-$S^2C^2$ outperforms the (10,7) MDS-coded computation approach by 17\%. As expected, (10,7)-$S^2C^2$ outperforms both (9,7) and (8,7)-$S^2C^2$ variants since the opportunities to do load balancing increase as the redundancy increases. The observed performance of the over-decomposition approach is lower than the performance of (10,7)-$S^2C^2$ owing to the extra data movement costs for load balancing during computations. Whereas in (10,7)-$S^2C^2$ there are no extra data movement costs during computations.

The wasted computation efforts measured in each of the worker node under (10,7)-coding are shown in figure \ref{fig:wastedWorkerWorst}. Due to a relatively high mis-prediction rate, $S^2C^2$ also incurs wasted computation among the worker nodes when the compute tasks of slow nodes are cancelled and reassigned to other worker nodes. However, the conservative $(10,7)$-MDS approach incurs higher wasted computation since it also ignores the slowest 3 nodes' computation efforts. On average, the conservative MDS scheme incurs 47\% more wasted computation effort.

\subsubsection{Results with $S^2C^2$ on polynomial coding}
\begin{figure}
    \centering
    \includegraphics[width=0.47\linewidth, height=4cm, trim=4 4 4 4, clip]{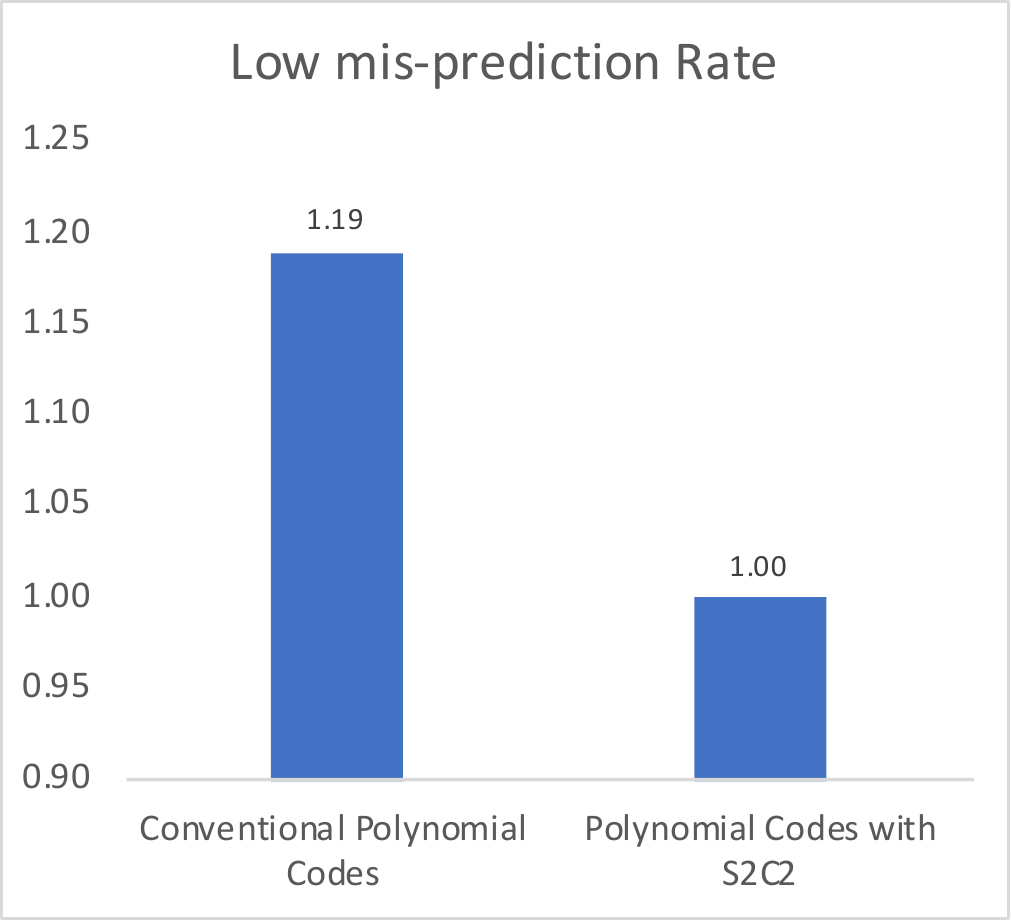}
    \includegraphics[width=0.47\linewidth, height=4cm, trim=4 4 4 4, clip]{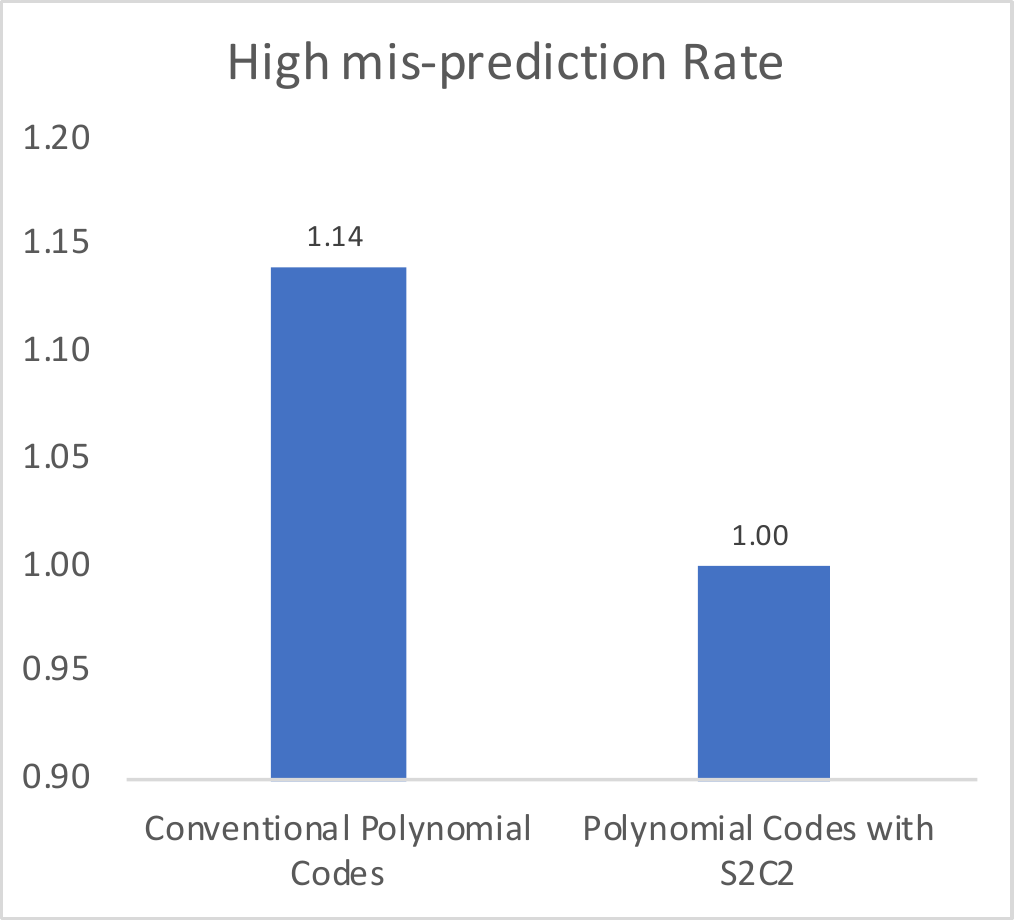}
    \caption{$S^2C^2$ on polynomial codes}
    \label{fig:perf_poly_no}
\end{figure}

We evaluate $S^2C^2$ applied on polynomial coding while performing Hessian matrix computation of the form $\textbf{A}^T diagonal(x) \textbf{A}$, as described in section \ref{subsec:LR_Ref}. The dimensions of matrix $\textbf{A}$ are $6000$ x $6000$. The results collected under low and high mis-prediction rates are shown in figure \ref{fig:perf_poly_no}. In these experiments, the cluster consists of 12 nodes. The matrices $\textbf{A}, \textbf{A}^T$ are partitioned each into 3 sub-matrices, encoded, and the encoded partitions are distributed to the 12 nodes. Each node would compute on 2 encoded partitions. Results from any 9 nodes would be enough to compute the Hessian. In this setup, $S^2C^2$ reduces the overall computation time by 19\% in low mis-prediction rate environment. The maximum possible reduction is $\frac{12-9}{9}=33.3\%$. The part of Hessian computation where each node has to first compute $diagonal(x)\tilde{A_i}$ is not influenced by $S^2C^2$. As a result, the gains from using $S^2C^2$ are lower than expected. Under high mis-prediction rate environment, $S^2C^2$ reduces the overall computation time by 14\%. 

\subsubsection{Scalability studies on a larger cluster}
\begin{figure}
    \centering
    \includegraphics[width=0.47\linewidth, height=4cm, trim=4 4 4 4, clip]{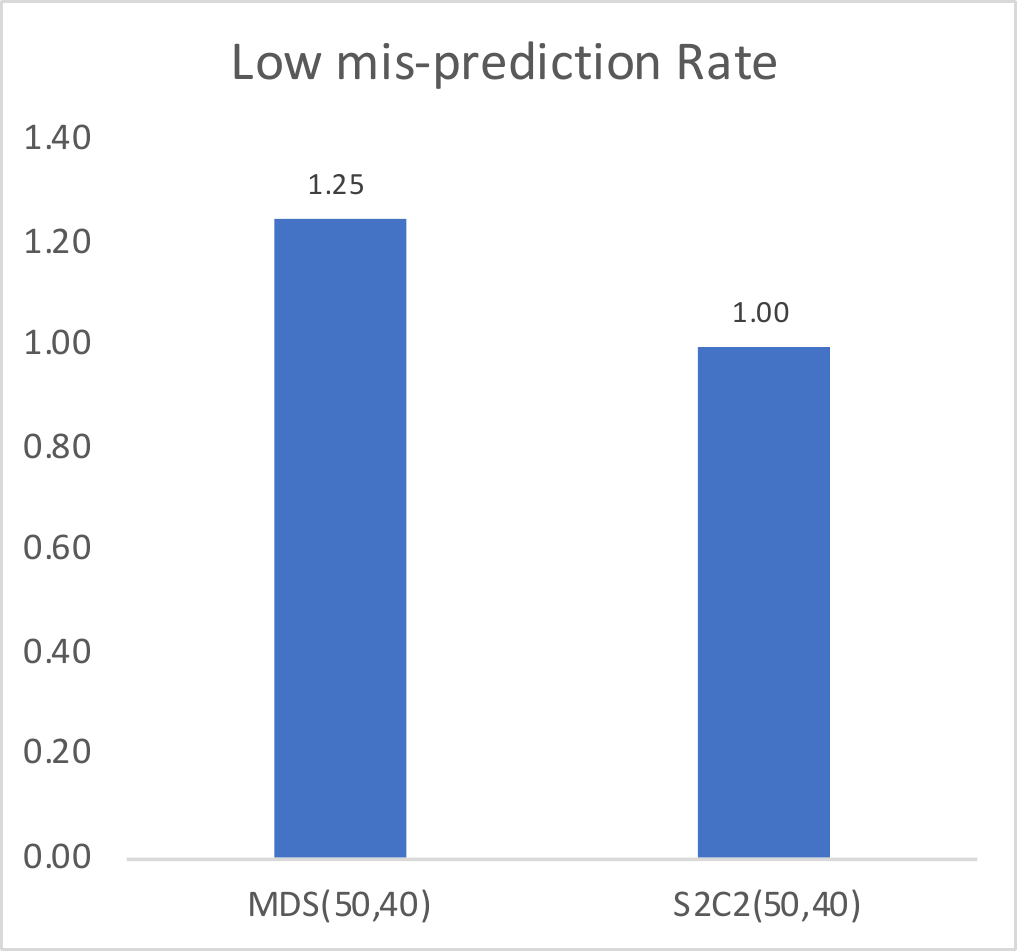}
    \includegraphics[width=0.47\linewidth, height=4cm, trim=4 4 4 4, clip]{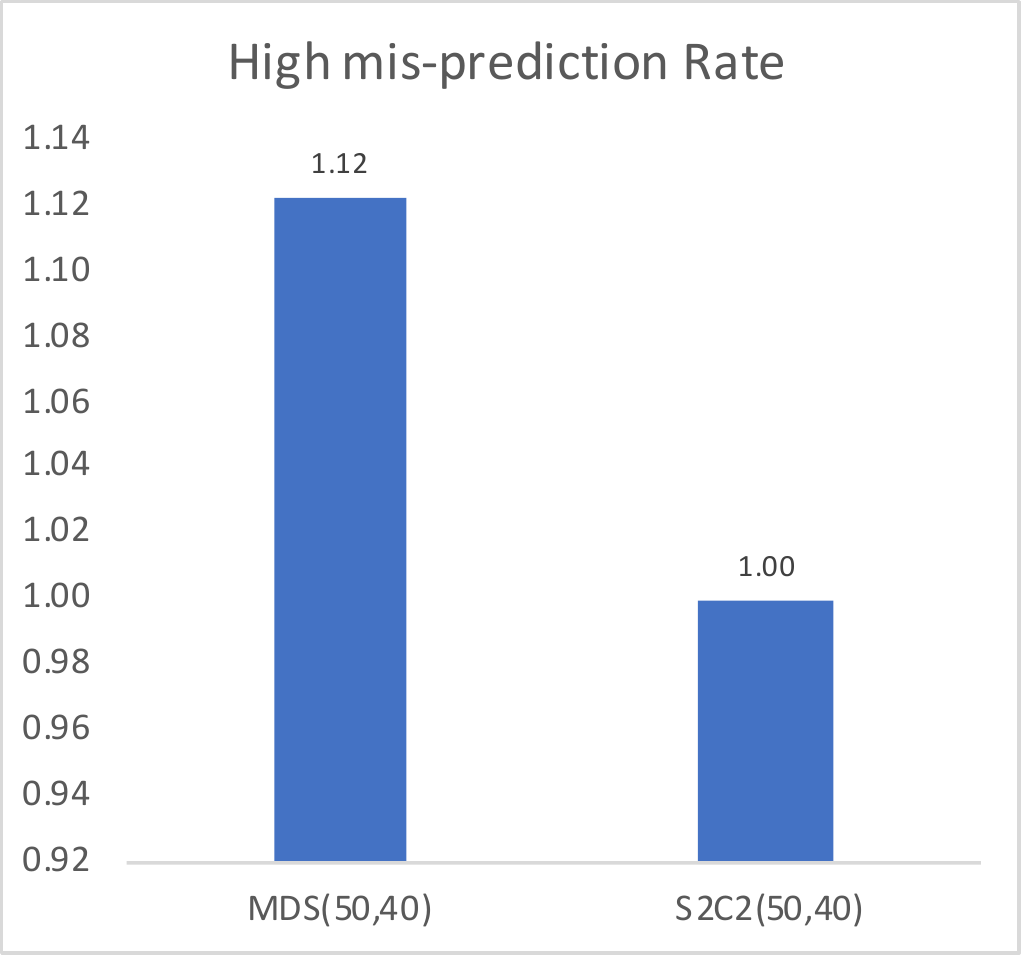}
    \caption{$S^2C^2$ performance on a 51 node cluster}
    \label{fig:perf_large_cluster}
\end{figure}

We performed experiments on a larger cluster with 50 worker nodes and one master node. Due to resource constraints we limit this scalability experiment to $S^2C^2$ and MDS coded computing approaches running SVM. We compared $S^2C^2$ and MDS coded computations under (50,40)-MDS codes while performing gradient descent for SVM. The results collected under low and high mis-prediction rates are shown in figure \ref{fig:perf_large_cluster}. For $S^2C^2$ the maximum reduction in execution time over (50,40)-MDS coded computation would occur when all 10 workers are always fast during execution. The exact reduction would be $\frac{50-40}{40}=25\%$. $S^2C^2$ reduces overall computation time by 25\% in a low speed mis-prediction rate environment. In a high mis-prediction rate environment, $S^2C^2$ reduces overall computation time by 12\%.

The evaluation results presented in this section demonstrate the effectiveness of $S^2C^2$ across different coded computation schemes and across different scales.

\section{Related Work}
\noindent \textbf{Straggler Mitigation}: There are several straggler mitigation techniques in literature which are reactive i.e., they wait until many tasks finish their execution before detecting and mitigating stragglers. The authors in  work~\cite{ananthanarayanan2010reining} utilize real time progress reports to detect and cancel the stragglers early. Authors in ~\cite{late} use LATE algorithm to improve the straggler detection and speculative execution in Hadoop framework. Adrenaline \cite{adrenaline} identifies and selectively speeds up long queries by quick voltage boosting. Authors of~\cite{tail} use software techniques such as selective replication of straggling requests. $S^2C^2$ differs from these techniques because it is a pro-active approach to straggler mitigation. In works \cite{tales, treadmill, kasture} the authors explore system sources of tail latency from system and implement mechanisms to eliminate these causes. In another set of works, \cite{quasar, dLo, jLeverich, compactorHarchol}, authors focus on improving resource efficiency while providing low latency. Both these works are complementary to $S^2C^2$ and can be used along with $S^2C^2$. Using replicated tasks to improve the response times has been explored in \cite{clones,nihar,wang,gardner,chaubey,lee}. This approach involves launching multiple copies of each task across workers, using results from the fastest copy and canceling the slower copies. This approach is pro-active like $S^2C^2$ but it needs multiple replicas of all the data resulting in large compute and storage overheads. $S^2C^2$, on the other hand, uses efficient coded replication and has significantly low overheads. Another strategy used for straggler mitigation is arriving at an approximate result without waiting on the stragglers \cite{approx,luo2019hop}. $S^2C^2$ does not do approximation and computes the precise result. \noindent

\textbf{Coded Computation}: Coded computation is a recently proposed framework with two concepts to deal with the communication and straggler bottlenecks in distributed computing. The first coded computing concept~\cite{LMA_all,li2016fundamental} enables an inverse-linear tradeoff between computation load and communication load in distributed computing. This can be leveraged to speed up large-scale data analytics applications~\cite{CTS16}. The second coded computation, the focus of this paper, concept~\cite{speedUpML} provides resiliency to stragglers and can be utilized to mitigate tail latency in distributed computing~\cite{speedUpML,reisizadehmobarakeh2017coded,LMA16_unify,dutta2016short,tandon2016gradient,polyCodes}. In particular, several of these works target distributed machine learning.There have been few recent works in the coded computing literature to exploit the computations of slow nodes~\cite{seqApp,YaoqingNIPS2017}, however the key ingredient of our proposed strategy is that it dynamically adapts the computation load of each node to its estimated speed from the previous rounds of computations.

\textbf{Performance Prediction}: Dinda et al. \cite{dinda_rta} described and evaluated, Running Time Advisor (RTA), a system that can predict the running time of compute-bound tasks. For predicting running time, linear time series analysis predictions of host loads are used.
Wolksi et al.  \cite{wolski_nws} developed Network Weather Service (NWS) to provide forecasts for network performance, and available cpu percentage at each compute node. NWS uses time series models, like ARIMA models, for forecasting. It maintains and updates multiple models, and dynamically selects the best performing model to provide forecasts. Speed prediction algorithm in $S^2C^2$ takes a similar approach to these prediction algorithms but uses the LSTM model to predict running times.

\section{Conclusion}
In this paper we proposed and evaluated $S^2C^2$ that efficiently tolerates speed variance and uncertainty about the number of stragglers in the system. $S^2C^2$ distributes coded data to nodes and during runtime adaptively adjusts the computation work per node. Thereby it significantly reduces the total execution time of several applications. Through our evaluations using machine learning and graph processing applications, we demonstrate \textasciitilde 39.3\% reduction in execution time in the best case. We conclude that speed adaptive workload scheduling as done by  $S^2C^2$   effectively reduce the overhead in coded computation frameworks and make them more effective in real deployments.

\section{Acknowledgements}
We sincerely thank all the reviewers for their time and constructive comments. We thank Daniel Wong for his valuable feedback on the paper manuscripts. We would also like to thank Saurav Prakash and Chien-Sheng Yang for the helpful discussions. This material is based upon work supported by Defense Advanced Research Projects Agency (DARPA) under Contract No. HR001117C0053, ARO award W911NF1810400, NSF grants CCF-1703575 and CCF-1763673, and ONR Award No. N00014-16-1-2189. The views, opinions, and/or findings expressed are those of the author(s) and should not be interpreted as representing the official views or policies of the Department of Defense or the U.S. Government.

\bibliographystyle{ACM-Reference-Format}
\bibliography{SC19_main.bib}

\end{document}